\PassOptionsToPackage{unicode}{hyperref}
\PassOptionsToPackage{hyphens}{url}
\documentclass[
  letterpaper,
  twocolumn,
  10pt]{article}
\usepackage{xcolor}
\usepackage{amsmath,amssymb}
\usepackage{iftex}
\ifPDFTeX
  \usepackage[T1]{fontenc}
  \usepackage[utf8]{inputenc}
  \usepackage{textcomp} %
\else %
  \usepackage{unicode-math} %
  \defaultfontfeatures{Scale=MatchLowercase}
  \defaultfontfeatures[\rmfamily]{Ligatures=TeX,Scale=1}
\fi
\usepackage{lmodern}
\ifPDFTeX\else
\fi
\IfFileExists{upquote.sty}{\usepackage{upquote}}{}
\IfFileExists{microtype.sty}{%
  \usepackage[]{microtype}
  \UseMicrotypeSet[protrusion]{basicmath} %
}{}
\makeatletter
\@ifundefined{KOMAClassName}{%
  \IfFileExists{parskip.sty}{%
    \usepackage{parskip}
  }{%
    \setlength{\parindent}{0pt}
    \setlength{\parskip}{6pt plus 2pt minus 1pt}}
}{%
  \KOMAoptions{parskip=half}}
\makeatother
\usepackage{color}
\usepackage{fancyvrb}

\DefineVerbatimEnvironment{Highlighting}{Verbatim}{commandchars=\\\{\}}
\usepackage{framed}
\definecolor{shadecolor}{RGB}{248,248,248}
\newenvironment{Shaded}{\begin{snugshade}}{\end{snugshade}}

\newcommand{\CommentTok}[1]{\textcolor[rgb]{0.56,0.35,0.01}{\textit{#1}}}

\newcommand{\NormalTok}[1]{#1}

\NewDocumentCommand\citeproctext{}{}
\NewDocumentCommand\citeproc{mm}{%
  \begingroup\def\citeproctext{#2}\cite{#1}\endgroup}
\makeatletter
 \let\@cite@ofmt\@firstofone
 \def\@biblabel#1{}
 \def\@cite#1#2{{#1\if@tempswa , #2\fi}}
\makeatother
\newlength{\cslhangindent}
\newlength{\csllabelwidth}
\newenvironment{CSLReferences}[2] %
 {\begin{list}{}{%
  \setlength{\itemindent}{0pt}
  \setlength{\leftmargin}{0pt}
  \setlength{\parsep}{0pt}
  \ifodd #1
   \setlength{\leftmargin}{\cslhangindent}
   \setlength{\itemindent}{-1\cslhangindent}
  \fi
  \setlength{\itemsep}{#2\baselineskip}}}
 {\end{list}}
\usepackage{calc}

\newcommand{\CSLLeftMargin}[1]{\parbox[t]{\csllabelwidth}{\strut#1\strut}}
\newcommand{\CSLRightInline}[1]{\parbox[t]{\linewidth - \csllabelwidth}{\strut#1\strut}}

\providecommand{\tightlist}{%
  \setlength{\itemsep}{0pt}\setlength{\parskip}{0pt}}
\PassOptionsToPackage{kerning,spacing}{microtype}
\usepackage{hyperref}
\usepackage[margin=1in]{geometry}
\usepackage{tikz}
\usepackage{filecontents}
\usepackage{booktabs, array}
\usepackage{dblfloatfix}
\usepackage{placeins}
\usepackage{makecell}
\usepackage{colortbl}
\usepackage{xurl}
\usepackage{tabularx}
\usepackage{listings}
\usepackage{chngcntr}
\usepackage{fvextra}
\DefineVerbatimEnvironment{Highlighting}{Verbatim}{ commandchars=\\\{\}, fontsize=\footnotesize, breaklines=true }
\hypersetup{colorlinks=true, linkcolor=blue, filecolor=magenta, urlcolor=cyan, citecolor=red}
\usepackage{bookmark}
\IfFileExists{xurl.sty}{\usepackage{xurl}}{} %
\hypersetup{
  pdftitle={S12X Patch Diffing with QBinDiff},
  hidelinks,
  pdfcreator={LaTeX via pandoc}}

\title{S12X Patch Diffing with QBinDiff}
\author{{\rm Ben Gardiner}\\
{\rm National Motor Freight Traf{}fic Association Inc.}\\
\textit{This is a preprint; final version presented at VehicleSec / Black Hat USA 2026}}
\date{}

\begin{document}
\maketitle
\begin{abstract}
This paper presents a reverse engineering analysis of a firmware update
for a commercial vehicle Brake ECU. We analyze the updater executable to
extract firmware and perform differential binary analysis of the S12X
architecture firmware images. Our research identifies specific changes
in the recall that address undocumented vulnerabilities in legacy
protocol processing. We demonstrate that the patched functionality
contained critical flaws. The findings confirm the safety recall
remediation was also a security patch.
\end{abstract}

\section{Introduction}\label{introduction}

Introduced circa 2001, the J2497 \citeproc{ref-sae2026j2497}{1} (aka
`PLC4TRUCKS') powerline databus remains the only industry-standard means
to satisfy the trailer ABS warning light requirement of FMVSS 121
\citeproc{ref-fmvss121}{2}, S5.1.6.2(b). As such, J2497 has been present
in all towing application Class 8 vehicles in North America from 2001 to
the present.

Previous research on J2497 ECU security \citeproc{ref-CVE-2020-14514}{3}
\citeproc{ref-gardiner2022disclosure}{4}
\citeproc{ref-nmfta2022actionable}{5} \citeproc{ref-CVE-2022-26131}{6}
revealed that J2497 is wirelessly reachable and trailer brake controller
ECUs implement diagnostics functions inside J1587 Data Link Escapes
beyond FMVSS 121 requirements. However, current threat models assume
tractor brake controllers only process necessary LAMP messages on J2497.

The 2024 Bendix EC80 safety recall
\citeproc{ref-bendix2024chronology}{7} revealed safety impacts to
tractor ECUs from J2497 reception, suggesting the EC80 processed J2497
traffic beyond necessary LAMP messages. The safety recall was remediated
via a firmware update of the EC80 using the `ID9363' updater executable.

We performed binary differential analysis on the S12X firmware (FW) pre-
and post-recall in 3 types of the EC80 (one for each OEM affected by the
recall). Results show extensive processing of J1587 messages received by
the tractor brake controller over J2497 in the removed code. This code
contained vulnerabilities including hardcoded secrets protecting
traction control configuration, buffer overflows, and memory corruption
leading to denial of service (of the vehicle) and remote code execution.
Finally, we identified the removal of PID handlers discoverable through
simple fuzzing. This suggests the safety recall also functions as a
security patch, preventing attackers with wireless, `adjacent'
\citeproc{ref-CVE-2022-26131}{6}, access to J2497 from exploiting these
vulnerabilities. This paper focuses exclusively on changes that were
`fixed' by this patch; no `0day' vulnerabilities are revealed as all
have been patched.

Contributions. In summary, we make the following contributions:

\begin{itemize}
\item
  \textbf{FW Update Process}: We reverse engineered the operations
  performed by ID9363 from several captures of the UDS
  \citeproc{ref-iso14229}{8} traffic (Section
  \ref{id9363-update-process})
\item
  \textbf{EC80 Firmware Partitions}: The flash layout/partitioning and
  fixed addresses of the EC80 are revealed within the constraints of the
  S12X architecture (Section \ref{ec80-bootloaders-and-application})
\item
  \textbf{S12X Reverse Engineering}: We present novel methods for
  reversing the S12X banked memory architecture. We detail memory map
  reconstruction, symbolic analysis, and opportunistic function
  identification, overcoming significant tooling gaps (Sections
  \ref{limitations-of-static-analysis},
  \ref{limitations-of-dynamic-analysis} and \ref{binary-diffing})
\item
  \textbf{The Security Patch}: The functionality removed by the patch is
  confirmed on all three types of EC80 units acquired and the
  vulnerabilities present in the removed functionality are itemized
  (Section \ref{changes-made-in-patch})
\item
  \textbf{Exploitability of Removed Functionality}: We evaluate the
  exploitability of each of the patched vulnerabilities from a mix of:
  bench and in-motion vehicle testing (Sections
  \ref{exploitability-of-removed-functionality} and
  \ref{notes-on-in-motion-vehicle-tests}).
\end{itemize}

\section{Background}\label{background}

In late 2024, three Class 8 vehicle OEMs in North America issued a
safety recall \citeproc{ref-paccar2024recall}{9}
\citeproc{ref-international2024recall}{10}
\citeproc{ref-volvo2024recall}{11}; these were the three OEMs that
integrate the Bendix EC80 brake controller. Bendix identified potential
memory corruption causing the ECU to go offline and worked with OEMs to
deploy a firmware update for affected trucks -- those with Electronic
Stability Program (ESP) or Automatic Traction Control (ATC) features --
estimated at 450,000 units at the time of this writing
\citeproc{ref-nhtsa_recalls_datahub}{12}. The timeline of this recall is
summarized below in Figure \ref{fig:timeline} (see also
\citeproc{ref-tsb10194446}{13} \citeproc{ref-tsb10176745}{14}
\citeproc{ref-tsb10222229}{15}). See \citeproc{ref-nmfta2025blog}{16}
for further details on the recall. The recall was later expanded in
October 2025 to cover some units which were sold as aftermarket
equipment \citeproc{ref-bendix2025expansion}{17}
\citeproc{ref-landline}{18} \footnote{all recalls for this same issue
  known at this time: 24V780000, 24V818000, 24V915000, 25E073000,
  25E077000, 25E078000.}.

\begin{figure}[htpb]
\includegraphics[width=\columnwidth]{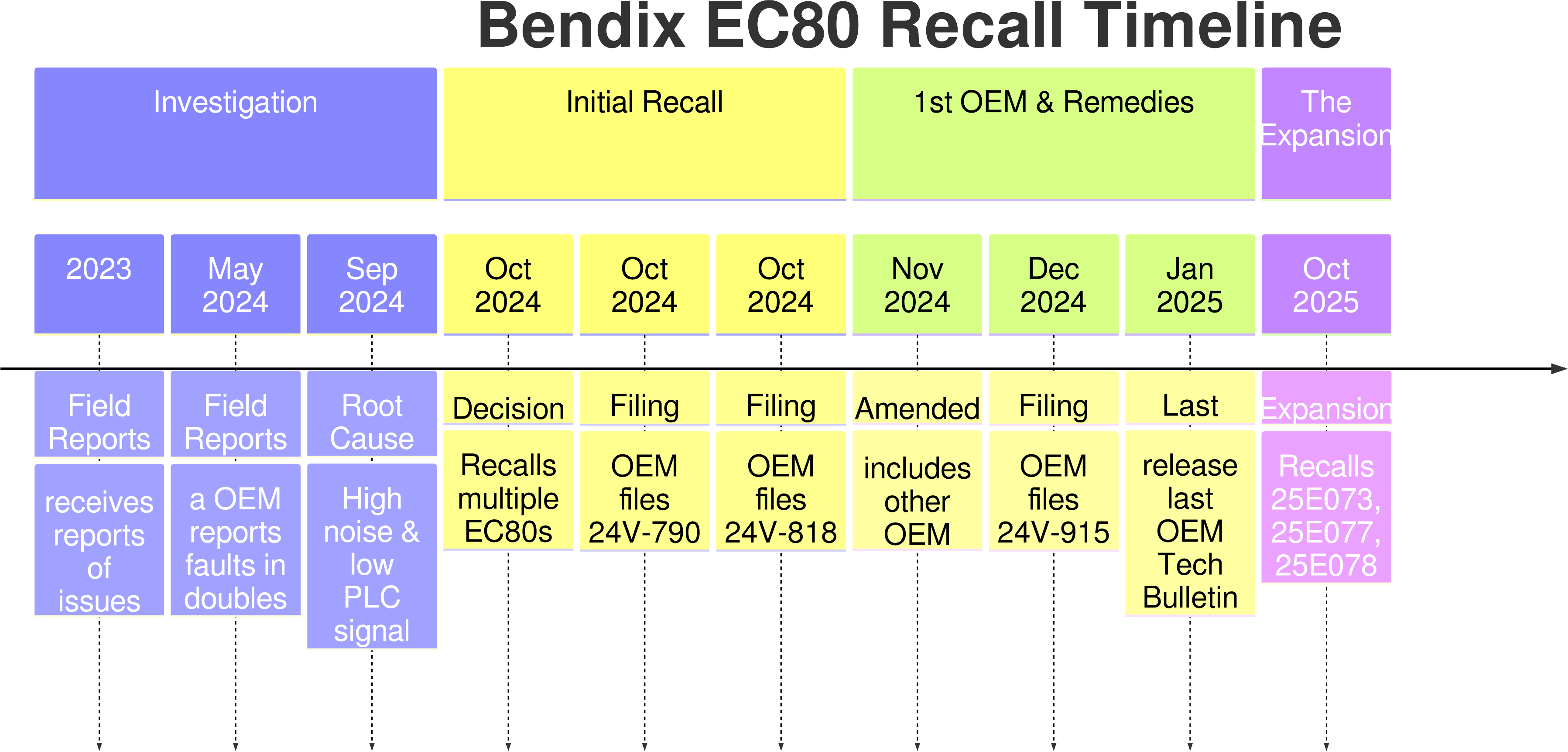}
\caption{Bendix Recall Timeline.}
\label{fig:timeline}
\end{figure}

The firmware update was distributed to fleets via a Windows executable,
`ID9363' \citeproc{ref-bendix2024volvo}{19}
\citeproc{ref-bendix2024international}{20}
\citeproc{ref-bendix2024paccar}{21}. We examine the ID9363 update
process and the resulting firmware changes in three EC80 ECUs (one per
OEM).

The firmware update, or `patch', purportedly prevents J2497 noise from
triggering memory corruption. However, due to CVE-2022-26131
\citeproc{ref-CVE-2022-26131}{6}, it is possible to wirelessly inject
signals onto J2497. It is also possible to inject from compromised
connected devices such as the increasingly common trailer telematics
devices. We examine if these memory corruptions could be triggered by
malicious actors using wireless (or wired) injection on J2497.

J1587 \citeproc{ref-saej1587}{22} serves as (roughly) the application
layer for J2497, while J1708 \citeproc{ref-saej1708}{23} defines the
original physical and data link layers (i.e.~``framing''). The J1708
physical layer is replaced with a Power Line Carrier (PLC) modulated
over the 12V auxiliary power line in J2497, but it inherits the J1708
data link and J1587 application layers. A frame can only be a maximum of
21 bytes long and must end with a one-byte checksum (the two's
complement of the sum of the preceding bytes). In software, the frames
are received byte-by-byte over a standard UART peripheral via a J1708 to
J2497 converter chip (the Intellon SSCP485 \citeproc{ref-sscp485}{24}).

For the J1587 application layer, the first byte is defined by J1708 and
is the Message Identification (MID), which identifies the sender or the
message type. MIDs 0x0A (LAMP ON), 0x0B (LAMP OFF), and 0x57 (Active
Trailer ABS Event) are required by the J2497 standard. For other J1587
messages the bytes following the MID are Parameter Identification (PID),
which specify the type of attached data (e.g., VIN, wheel speed,
diagnostics) followed by parameters (payload). Some PIDs have a fixed
data length (like a 1-byte speed value), while others are variable
length. PIDs data enables transmission of vehicle signals, diagnostic
commands, and proprietary data.

The Bendix EC80 is a heavy-duty vehicle Electronic Control Unit (ECU)
responsible for Anti-lock Braking (ABS), ATC, and ESP functions. It
controls pneumatic circuits through external pressure modulation valves
connected to PCB-mounted FET drivers. During initialization, absent
critical Diagnostic Trouble Codes (DTCs), modulators undergo individual
testing, producing a `roll call' of small air chuffs \footnote{In a
  bench environment the same successful power-on results in a clicking
  of a failsafe `diagonal' relay.}.

Table \ref{tab:ec80targets} provides an overview of the EC80 units
acquired for this study. The complexity of the EC80 firmware varies with
the integrated features. The middle target, \texttt{2ec80}, represents a
medium complexity unit and was selected as the primary target for
reverse engineering in this paper. All findings have been validated
across all the target firmwares.

\begin{table}[htbp]
\centering
\footnotesize
\begin{tabular}{@{} l p{0.22\columnwidth} p{0.22\columnwidth} p{0.22\columnwidth} @{}}
\toprule
Target Name & \texttt{1ec80} & \texttt{2ec80} & \texttt{3ec80} \\
\midrule
FW version, before & Z228999 & Z266494 & Z286098 \\
FW version, after & Z300822 & Z302578 & Z302579 \\
Sticker Name & EC80ESP+ & EC80ESP & EC80ESP \\
Non-sticker Feature & J1708 & - & - \\
Sticker Features & 6S/6M & 6S/6M & 4S/4M \\
 & PLC & PLC & PLC \\
 & 2nd CAN & 2nd CAN & - \\
 & - & CAN Gateway & - \\
 & Integrated TPMS & - & - \\
HW Date Code & 20220723 & 20230507 & 20221019 \\
SW Date Code & 3H0922G & 3E2323G & 2F1423G \\
\bottomrule
\end{tabular}
\caption{Information on the three EC80s acquired and tested. While specific part numbers and OEM integrators are omitted, the firmware versions presented are consistent with recall documentation and can be retrieved from the target ECU via a Read Data By Identifier request (0xF194). The order here matches Figure \ref{fig:pcbs} left to right.}
\label{tab:ec80targets}
\end{table}

The EC80 units affected were released in April 2020
\citeproc{ref-bendix2024international}{20}. Given its role in preventing
rollovers and maintaining stability, brake controllers are expected to
have a high Automotive Safety Integrity Level (ASIL), likely ASIL C or
D, to meet OEM safety targets. While specific ASIL targets for EC80
vehicles are unpublished, Hazard Analysis and Risk Assessment (HARA) of
brake controllers \citeproc{ref-funcbrakearch}{25} identified ASIL D
scenarios. While focused on autonomous systems, this analysis notes the
architecture applies to non-autonomous vehicles, stating the brake
system is critical regardless of autonomy. Furthermore, the literature
for a competing product, the mBSP, explicitly states compliance with ISO
26262 \citeproc{ref-iso26262}{26} up to ASIL D
\citeproc{ref-zfmbsp}{27}.

A Threat and Risk Assessment (TARA) of Class 8 vehicles
\citeproc{ref-nmfta2024requirements}{28} ranked brake controllers third,
after telematics and gateway devices, for cybersecurity priority. The
EC80 connects to multiple interfaces, including ``Untrustworthy Network
Domains'' like J2497 \citeproc{ref-gardiner2024security}{29}. Because it
is not explicitly intended to transport, translate, or filter traffic,
it is classified as an ``unintended gateway'' device
\citeproc{ref-gardiner2024security}{29} \footnote{Although the `Z'
  version which implements a `CAN Gateway' (see Table
  \ref{tab:ec80targets}) could be considered a `(intentional) gateway
  device'}. Consequently, it must satisfy cybersecurity requirements
NGW-S-001 (Security Assurance), NGW-S-002 through -005 (Won't Transport,
Translate, Filter, etc.) \citeproc{ref-gardiner2024security}{29}.

\begin{figure}[htpb]
\includegraphics[width=\columnwidth]{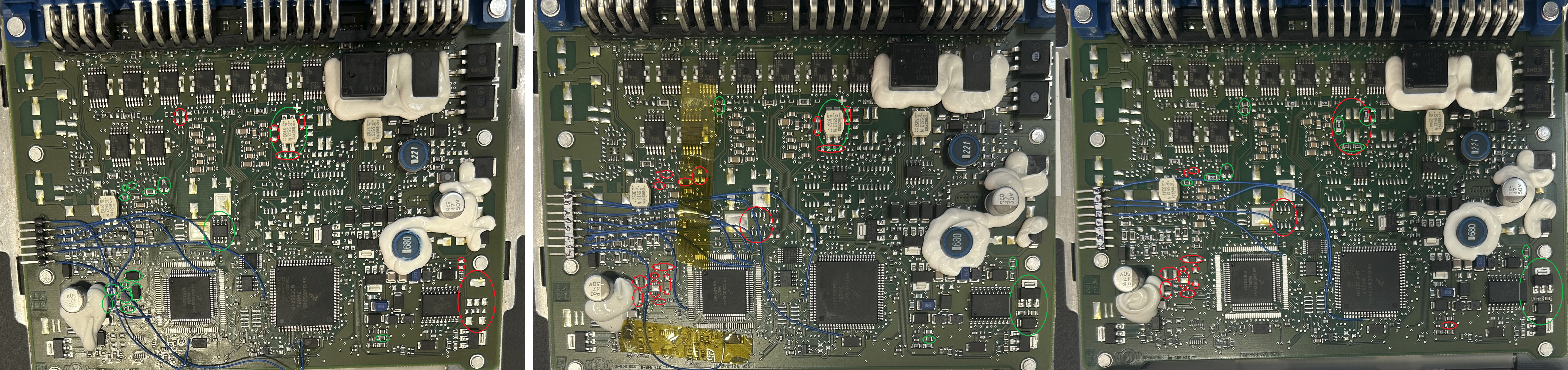}
\caption{The PCBs of the three EC80s acquired and tested. A 0.1" header is installed for access to the BDM programming pins on the right S12X MCU. The added/removed components are highlighted in green/red, respectively.}
\label{fig:pcbs}
\end{figure}

The EC80 uses two S12X microcontrollers (as can be seen in the PCBs
pictured in Figure \ref{fig:pcbs}). Both are specifically the NXP
MC9S12XEQ512 \citeproc{ref-mc9s12xeq512}{30}, a 16-bit microcontroller
from the S12X family; however, the `left' MCU is in a smaller footprint
80-pin QFP package compared to the larger `right' MCU in a 144-pin LQFP
package.

The S12X \citeproc{ref-s12xuserguide}{31} features an XGATE peripheral
coprocessor to offload high-speed data transfer and logic processing
from the main CPU12X core. The S12X employs a paging architecture using
PPAGE, EPAGE, and RPAGE registers to map 16-bit logical windows into a
23-bit global address space. The device includes 512 KiB of PFLASH
(Program Flash) residing at global addresses 0x780000 - 0x7FFFFF and 32
KiB of DFLASH (Data Flash) located at 0x100000 - 0x107FFF. Additionally,
it features a 4 KiB Emulated EEPROM (EEE) buffer RAM mapped to global
addresses 0x13F000 - 0x13FFFF, which allows the firmware to treat the
DFLASH as random-access non-volatile memory.

\section{ID9363 Update Process}\label{id9363-update-process}

\begin{figure*}[!b]
\centering
\includegraphics[width=\textwidth]{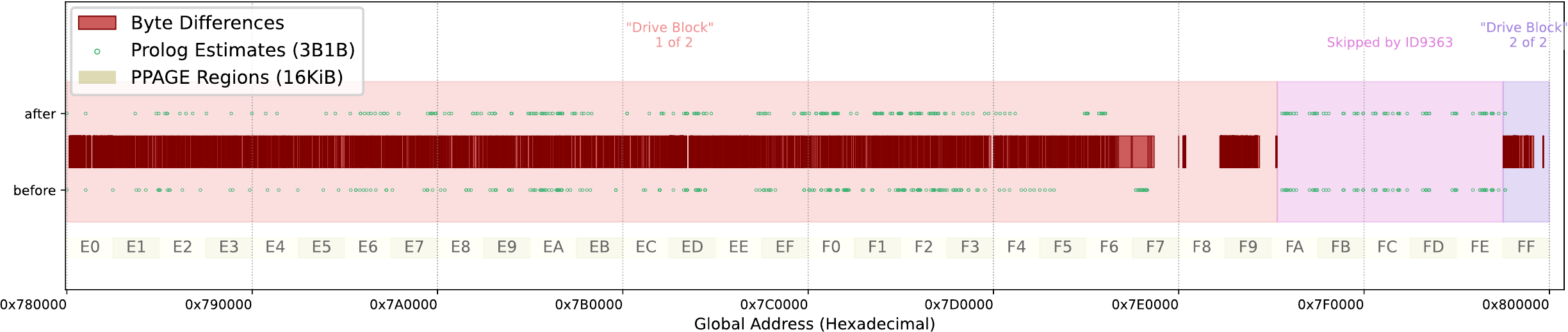}
\caption{Comparison of the PFLASH of 2ec80 before and after ID9363 update. The unchanged extents illustrate that both the interrupt vector table at the end of PFLASH (0x7E8800) and the region 0x7E8800 - 0x7FC000 were skipped by the firmware update. Many other locations where the PFLASH contents are unchanged were found to have value 0x3F '?', the Software Interrupt (`SWI`) instruction, which we believe is used as a padding value. The number of 0x3F bytes increased, suggesting function deletion by the update. Function prolog estimate locations across the update suggest shifting to lower addresses which (also) suggests function deletion by the update (examined in Section \ref{changes-made-in-patch}).}
\label{fig:bytediffs}
\end{figure*}

We observed several EC80 firmware updates. The new firmware is readable
in cleartext from CAN databus captures; however, analyzing the update
requires the pre-update firmware contents. While some ECUs offer a UDS
`upload' service, we confirmed the EC80 does not. From the traffic we
reconstructed the seed-key exchange routine in closed-form for all
DSC,SA sessions (see Appendix
\ref{id9363-update-process-further-details}), confirming no firmware
upload service exists in the observed sessions (DSC=2,SA=5) (DSC=3,SA=1)
(DSC=3,SA=7) (DSC=3,SA=3) \footnote{memory leaks, other unintended
  upload functionality in the bootloader or application are out of the
  scope of this paper}.

We used the BDM interface to dump PFLASH, DFLASH, and EEE (emulated in
DFLASH) from both MCUs to obtain the pre-update firmware. We found that
both the XPROG tool \citeproc{ref-xprog}{32} and PROGS12Z
\citeproc{ref-progs12z}{33} worked well. XPROG creates flat binary files
of PFLASH, DFLASH and EEE which correspond to their global address
extents. PROGS12Z creates .s19 files \citeproc{ref-s19}{34} which
contain the global address extents of PFLASH, DFLASH, and EEE.

We found that although the S12X used in the EC80 supports read-out
protection, it was not enabled. We confirmed that the data transferred
over CAN matches the PFLASH contents after update. Only the `right' MCU
(larger package) receives code changes; the `left' MCU's PFLASH remained
unchanged. It did, however, receive some DFLASH / EEE changes during the
update process. We focus solely on the `right' MCU's code changes.

A visualization of the byte-by-byte difference of PFLASH before and
after update of a \texttt{2ec80} is presented in Figure
\ref{fig:bytediffs}. Byte differences are marked by pink fill, red
outline boxes, and are numerous (this is clearly not a micropatch
\citeproc{ref-ampprogrammicropatching}{35}). The ``Drive Block''s 1 and
2 are labeled. The PPAGE numbers corresponding to the PFLASH offset are
labeled at the bottom for reference. The same style of visualization
will be reused throughout the paper to illustrate successive analysis
steps. In this figure function prolog estimates are plotted as green
circles (in subsequent figures the functions resulting from firmware
analysis will replace these).

Despite extensive byte differences, the high-level source code changes
may be concise. Small machine code changes and linking order can cause
massive byte differences. Byte-level analysis cannot reveal specific
code changes or isolate areas of interest, as changes span `Drive
Blocks' 1 and 2. Understanding the changes requires disassembling and
comparing the pre- and post-update images.

\section{EC80 Bootloaders and
Application}\label{ec80-bootloaders-and-application}

Global PFLASH byte differences (Figure \ref{fig:bytediffs}) and ID9363
updater traffic analysis (Table \ref{tab:update_steps}) indicate a J1939
\citeproc{ref-saej1939}{36} CAN UDS \citeproc{ref-iso14229}{8}
bootloader in the skipped region. We label this the `late' bootloader as
it does not contain the reset vector. Code between the reset vector and
this late bootloader is labeled the `early' bootloader. Its code likely
resides in some subset of PPAGE 0xFF (aka ``Drive Block 2 of 2'') which
has a fixed mapping in the S12X memory space.

The bootloaders presumably hand-off to the `application'. Its code is
very likely in the ``Drive Block'' 1 of 2 region, although it could have
some code in the other ``Drive Block'' as well to make use of the fixed
mapping of PPAGE 0xFF.

\section{Limitations of Static
Analysis}\label{limitations-of-static-analysis}

\begin{figure*}[!b]
\centering
\includegraphics[width=\textwidth]{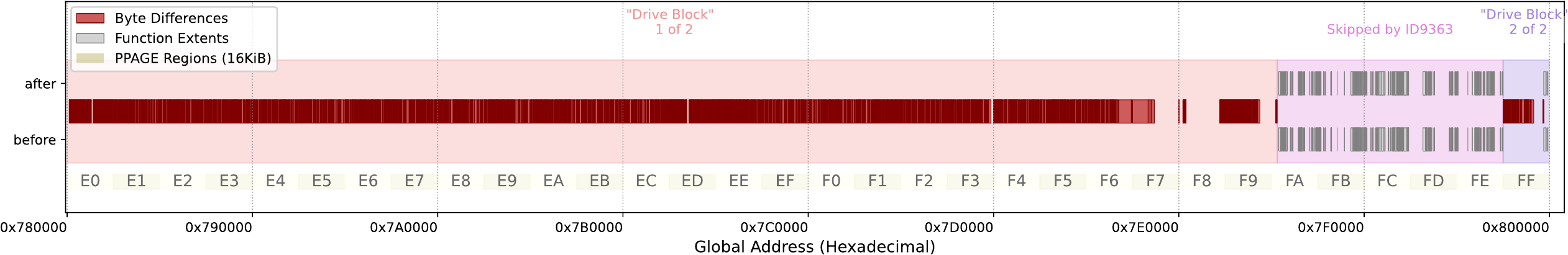}
\caption{Visualization of what the IDA Pro auto-analysis achieved after creating entrypoints for each of the handlers in the default interrupt vector table. The lack of functions in all but the 'early' and 'late bootloader' regions illustrates that auto-analysis is not able to follow the exeution flow into the 'application' region.}
\label{fig:demo_ivbr0xff}
\end{figure*}

\begin{figure}[htpb]
\includegraphics[width=\columnwidth]{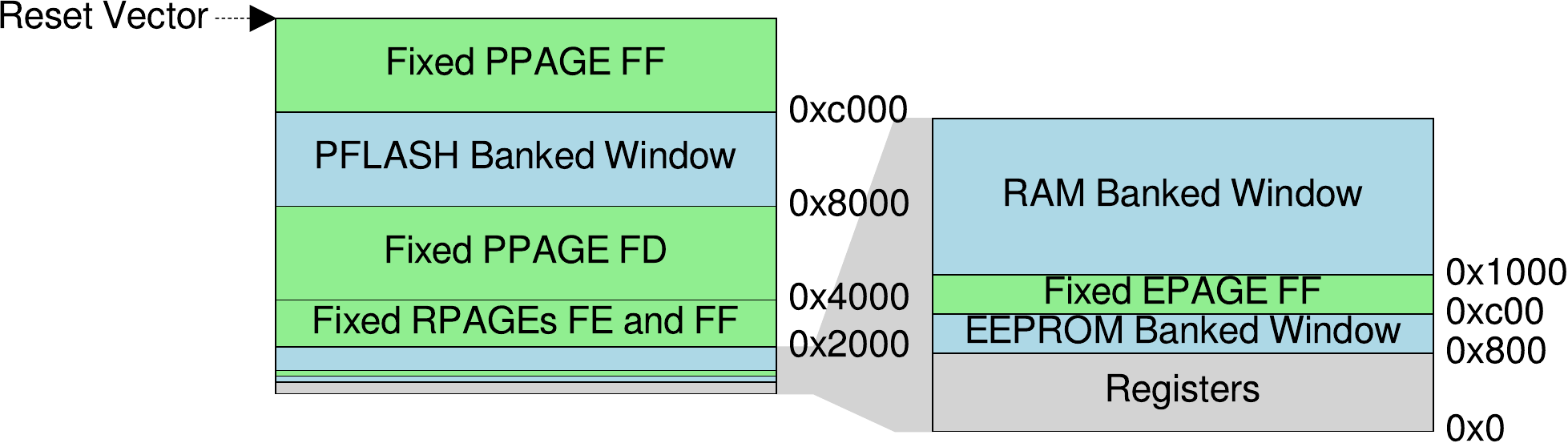}
\caption{Local 16-bit memory map of S12X by linkerscope. Green for fixed page areas and blue for paged windows.}
\label{fig:ppage-linkerscope}
\end{figure}

The S12X primarily uses a 16-bit memory space with banked windows
(PPAGE, DPAGE, RPAGE) to access global memory
\citeproc{ref-mc9s12xeq512}{30}, rather than direct global addressing
(Figure \ref{fig:ppage-linkerscope}) \footnote{although specific, slower
  execution, instructions exist for direct global address access.}.
Disassemblers must model this mapping to correctly resolve
page-dependent references to the global address space. Fixed mappings,
like PPAGE 0xFF at 0xC000-0xFFFF, are straightforward. The reset vector
is in this address so starting analysis with a disassembler is likewise
straightforward. In all EC80 firmwares, execution quickly encounters a
call instruction with a 3-byte target: a 16-bit offset and a PPAGE. At
this point the disassembler needs to have the code in the correct
location in its memory model.

\begin{figure}[htpb]
\includegraphics[width=\columnwidth]{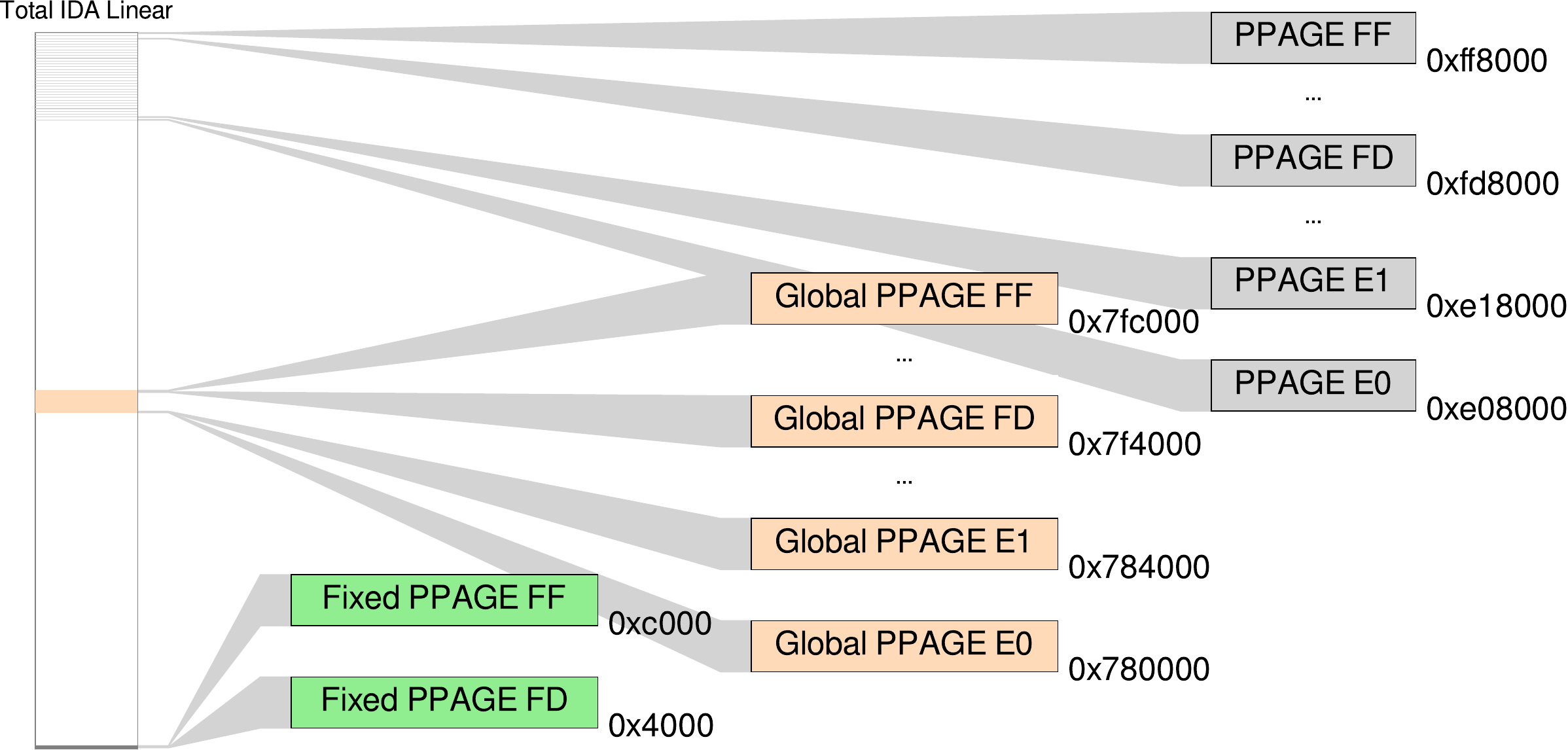}
\caption{Illustration of the copies of PFLASH placed in the IDA Linear Address space memory map for some success in static analysis.}
\label{fig:ida-linkerscope}
\end{figure}

There are not many examples of S12X architecture reversing available. Of
course the seminal work of \citeproc{ref-miller2013adventures}{37} did
include S12X reversing of a power assist steering module, and it did
include uncovering the use of hardcoded passwords there for UDS
\citeproc{ref-iso14229}{8} seed-key exchange (the security access \$27
service). But the `how' of creating dumps of PFLASH, DFLASH, RAM and
assembling it for a reverse engineering tool (e.g.~IDA Pro as used by
\citeproc{ref-miller2013adventures}{37}) are not available. The recent
analysis by Pulse Security of a motorcycle ECU
\citeproc{ref-pulsesecurityducati}{38} demonstrates a way to assemble an
S12X target binary image for analysis in Ghidra
\citeproc{ref-ghidra}{39} before switching to hunting for numeric tables
in Flash. Ghidra does not support the XGATE coprocessor, whereas IDA Pro
\citeproc{ref-idapro}{40} does; therefore IDA Pro was selected for this
analysis. The IDA Pro support team was kind enough to provide a code
snippet of how PPAGE is modeled in the IDA linear address space model.
It corresponds to how all of Ghidra, the HSW12 open source assembler
\citeproc{ref-hsw12}{41}, and the original S12X debugger, HiWave
\citeproc{ref-hiwave}{42} treat it: use PPAGE as the most significant
byte in a three-byte address (e.g.~the gray boxes in Figure
\ref{fig:ida-linkerscope}). The same is done for RPAGE and EPAGE
\footnote{This causes IDA linear space collisions between EPAGE
  0x10/0x13 and DFLASH global addresses. No workaround exists; users
  must be cautious if firmware uses both.}. Because RPAGE and PPAGE
windows are discontiguous in the 16-bit local space this does not cause
collisions. This addressing scheme is very useful and will be used in
the remainder of the paper (i.e.~any three byte addresses other than
global addresses imply the first byte is the page and the remaining
bytes are the offset within it). IDA also expects to be able to address
the global memory space in the exact same address values in its linear
memory map (the peach boxes in Figure \ref{fig:ida-linkerscope});
however, this is only used for instructions that do global memory space
access.

We used IDA Pro Python scripts to map PFLASH to global addresses and
copy it to high-byte and fixed PPAGE locations in the 16-bit local space
(Figure \ref{fig:ida-linkerscope}, Appendix \ref{ida-pflash-setup}). We
then created an entrypoint for every interrupt vector in 0xFF80 -
0xFFFF, including reset, CAN, Programmable Interrupt Timer (PIT), Serial
Communications Interface (SCI), and all other peripherals with an
interrupt. IDA Pro analysis results are shown in Figure
\ref{fig:demo_ivbr0xff}, with function extents replacing the prolog
estimates used previously in Figure \ref{fig:bytediffs}.

\begin{figure*}[htpb]
\centering
\includegraphics[width=\textwidth]{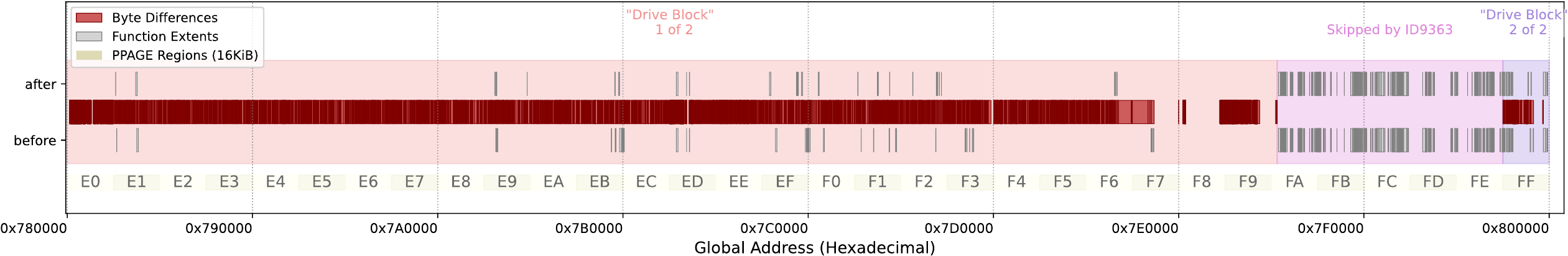}
\caption{Visualization of what the IDA Pro auto-analysis achieved after creating entrypoints for each of the handlers in an interrupt vector table given by IVBR=0xF7. Compared with Figure \ref{fig:demo_ivbr0xff}, this figure shows several functions in the 'application' region; however, they represent a small fraction of the firmware's total application code, the upper bound of which is given by the byte differences. Similar to the prologs in Figure \ref{fig:bytediffs}, it also shows correlated, left-shifted function extents which (again) suggest code deletion by the update; however, the low overall coverage limits our confidence.}
\label{fig:demo_ivbr0xf7}
\end{figure*}

Analysis proceeds past three-byte calls, confirming correct PPAGE
installation. While analysis reaches the late bootloader, it fails to
reach application code. Starting from the reset vector often yields
incomplete coverage due to unresolved data-dependent jump or call
targets in many firmwares and the EC80 is no exception.

We focus on J2497 processing changes, so understanding the entire EC80
bootloader is unnecessary. The EC80 receives J2497 data via an Intellon
SSCP485 \citeproc{ref-sscp485}{24} chip connected to the S12X SCI2
peripheral. However, the SCI2 interrupt vector of the default table
points to a dummy function, and no discovered functions reference SCI2
data registers.

\section{Limitations of Dynamic
Analysis}\label{limitations-of-dynamic-analysis}

Since static analysis failed to reveal SCI2 access, we used the unlocked
BDM for dynamic analysis.

The debugger software we had access to is HiWave
\citeproc{ref-hiwave}{42}. Breakpoints on SCI2 register reads were never
triggered. We found that breakpoints cease functioning after the late
bootloader entrypoint (0xFBADB1). The cause is unknown; possibilities
include \texttt{BGND} instruction execution or resets triggered by the
secondary (`left') processor. Attempts to reset both MCUs in tandem
failed. Despite evidence of debug-aware firmware (e.g.~watchdog feeding
after \texttt{STOP}), we abandoned further investigation to focus on the
firmware patch.

BDM allows reading S12X address spaces without interrupting execution
(mostly \footnote{BDM accesses global address space bypassing the MPU.
  It is non-intrusive only during free bus cycles, otherwise stealing
  cycles or stalling the CPU.}). We used HiWave's
\citeproc{ref-hiwave}{42} scripting interface (Appendix
\ref{hiwave-script}) to create hexdumps and convert them to .s19 files.
This allowed us to read `live' RAM and register contents without
breakpoints and load them into IDA.

Polling registers revealed that the application uses an IVBR of 0xF7,
unlike the initial bootloader's default 0xFF. The S12X has the
capability to switch the location of its interrupt vector table
\footnote{Except for Power-on, Clock Monitor, and COP Watchdog reset
  vectors.}; although static analysis at the stage shown in Figure
\ref{fig:demo_ivbr0xff} missed the IVBR assignment, we had confirmed the
application uses an interrupt vector table at 0xF710-0xF7F9.

We updated the IDA Pro database with RAM dumps and entrypoints for the
new vectors (Figure \ref{fig:demo_ivbr0xf7}). Analysis still covers much
of the late bootloader, as the application frequently calls back into it
(e.g., for programming mode).

We can now access application interrupt handlers. Target \texttt{2ec80}
supports XGATE \footnote{We analyzed XGATE code by mapping PFLASH and
  RAM in IDA. XGATE handled only ECT interrupts (channels 0-7) and was
  irrelevant to J2497 reception, so analysis ceased. Ghidra would
  suffice here, but we had at this point invested in IDA-based
  automation.}, PIT, CAN0, CAN4, Enhanced Capture Timer (ECT), SPI1, and
SCI2. We focus solely on SCI2. The SCI2 ISR, \texttt{sub\_C12B}, calls
functions where IDA Pro analysis fails due to lack of `register
tracking' for S12X. For now, we manually resolved data-dependent
\texttt{call} instructions (e.g., \texttt{call\ {[}-\$2662,y{]}}) that
IDA failed to handle. We also manually resolved jump tables, which IDA
Pro does not support on S12X.

Then IDA Pro analysis revealed the application's J2497 message reception
operation \footnote{One firmware change is observable now but will be
  detailed in Section \ref{changes-made-in-patch}.}. The details of the
receiver architecture are captured in Appendix
\ref{j2497-reception-architecture-of-ec80}; but the dataflow of received
packets from the SCI2 peripheral to consumers is simpler: frames are
built by appending to a receive buffer at 0x3BF5 and frame reception is
signaled with the frame counter / semaphore at 0x3BF4. Figure
\ref{fig:sci2_dataflow} illustrates this dataflow. This IDA analysis
identified only the `publish' side. No consumers of the frame counter or
receive buffer were found. Firmware typically splits interrupt handling
into a lightweight top-half and a main-thread bottom-half to minimize
interrupt context cycles.

\section{Binary Diffing}\label{binary-diffing}

Interrupt context execution was explored, but the main execution thread
remained invisible. Analysis failed to trace the late bootloader
hand-off to the application \footnote{Presumably due to at least a
  data-dependent call target but possibly due to this and other reasons.
  The result was good enough to identify the functions that consume the
  receive buffer 0x3BF5 and Frame Counter 0x3BF4: \texttt{sub\_EEB98}
  and \texttt{sub\_F096F9}; these are functions that check for LAMP
  messages and then process non-LAMP J1587 messages, respectively.}.
Searching for function prolog byte sequences yielded excessive false
positives. We observed function pointer arrays in `Drive Blocks' 1 and
2. Indirect call instructions (e.g., \texttt{call\ {[}-\$2662,y{]}})
read a 3-byte address (PPAGE offset + PPAGE value) from a resolved
memory location. The compiler/linker used here always emits a trailing
zero byte. Scanning for these patterns, filtering for valid PPAGE values
and offsets (0x8000-0xC000), significantly reduced false positives. We
further filtered out isolated function pointers, as they typically occur
in arrays. See Appendix \ref{ida-func-ptr-discovery} for the script
used. Then analysis discovered the main-thread bottom-half function for
J2497 message processing at sub\_EEB989 (see Figure
\ref{fig:sci2_dataflow2}).

The main thread processing in \texttt{sub\_EEB989}, handles J2497 LAMP
ON (0x0A), LAMP OFF (0x0B), and Active Trailer ABS Event (0x57)
commands, ignoring payloads. It consumes contiguous 0x0A, 0x0B, or 0x57
messages. Messages with other MIDs are passed to \texttt{sub\_F096F9}.
J1587 messages can contain one or more PID payloads. This function
splits J1587 messages into PID payloads and passes the payloads to
\texttt{sub\_F2A3E1}, which dispatches each to a handler from PFLASH
tables starting at the 0xDC3D `context structure' (see Appendix
\ref{j1587-pid-processing} for details). This data flow processing is
illustrated in Figure \ref{fig:sci2_dataflow2}.

To analyze six firmware images (three pairs) with BinDiff
\citeproc{ref-bindiff}{43}, we automated IDB creation (see
\ref{ida-pflash-setup}), segment setup, interrupt vector definition, and
function search. We scripted the identification and marking of PID
tables and handlers (Appendix \ref{ida-pid-naming}). We excluded the
late bootloader to focus BinDiff on the application firmware.

Lack of `register tracking' hindered automatic analysis, requiring
manual resolution of indirect call targets. We implemented rudimentary
concolic analysis to parse disassembly and symbolically track execution,
substituting concrete values from PFLASH, DFLASH, or RAM when needed and
where available. See Appendix \ref{ida-concolic-analysis} for the script
used. We also implemented basic jump table detection and creation by
parsing disassembly and creating manual code xrefs (Appendix
\ref{ida-jump-table}) \footnote{because the actual jump table operation
  ``probably can't be scripted'' according to IDA Pro support.}.

\begin{figure}[htpb]
\includegraphics[width=\columnwidth]{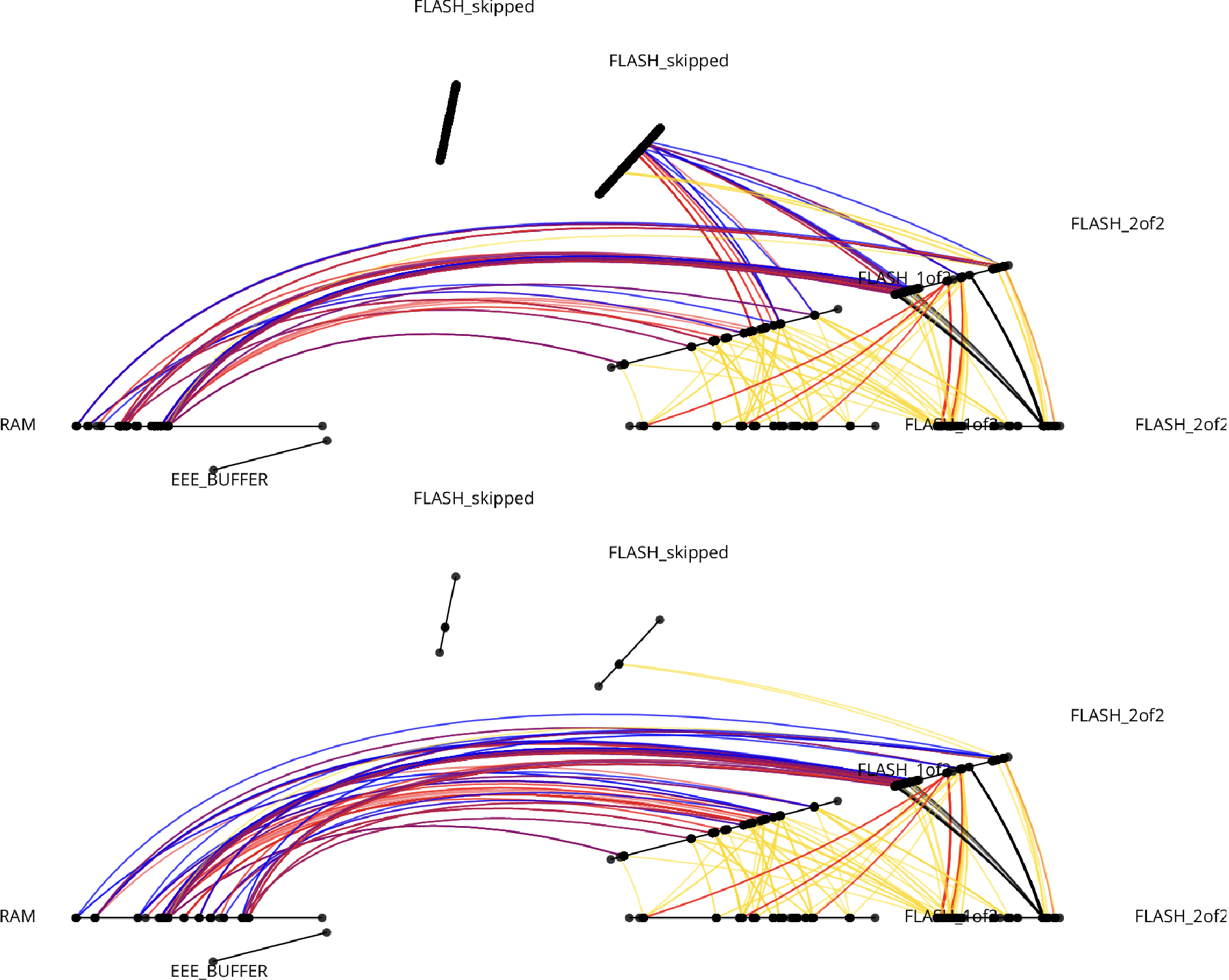}
\caption{Hiveplot panel comparing IDA Pro analysis achieved before (above) and after (below) a change of the 0x4000 mapping from Flash to RAM. The cross-references (xrefs) are drawn as edges directed counter-clockwise (CCW); each region with executable code is doubled to show intra-region xrefs; the "Drive blocks" 1 and 2 are shown co-linearly and the skipped region is shown lifted above them; yellow edges are code xrefs, red edges are data reads, blue are data writes and black are interrupt vector references.}
\label{fig:hiveplot-compare}
\end{figure}

To ensure BinDiff fidelity, we aimed for zero disassembly errors. We
found numerous disassembler errors in PPAGE 0xFD functions. Hiveplots
\citeproc{ref-krzywinski2011hiveplots}{44} of IDA Pro xrefs revealed
invalid data writes to PPAGE 0xFD from functions in other PPAGEs. These
writes are invalid as PFLASH requires sector erasure. Runtime polling of
the MMCCTL1 register confirmed that the application maps 0x4000 to fixed
RAM, not PPAGE 0xFD. We adjusted the IDA database segment setup, as
shown in Figure \ref{fig:hiveplot-compare}.

These steps achieved function coverage of 80\% (1ec80), 60\% (2ec80),
and 75\% (3ec80) of byte differences in PPAGEs E0-F9. We then
automatically performed BinDiff on all three target pairs. Bindiff was
able to support the S12X architecture and give reasonable insight;
however, its shortcomings when applied to this firmware caused spurious
matches e.g.~it matched the SCI2 handler in a before image with the SCI3
handler in an after image.

\begin{figure*}[htpb]
\centering
\includegraphics[width=\textwidth]{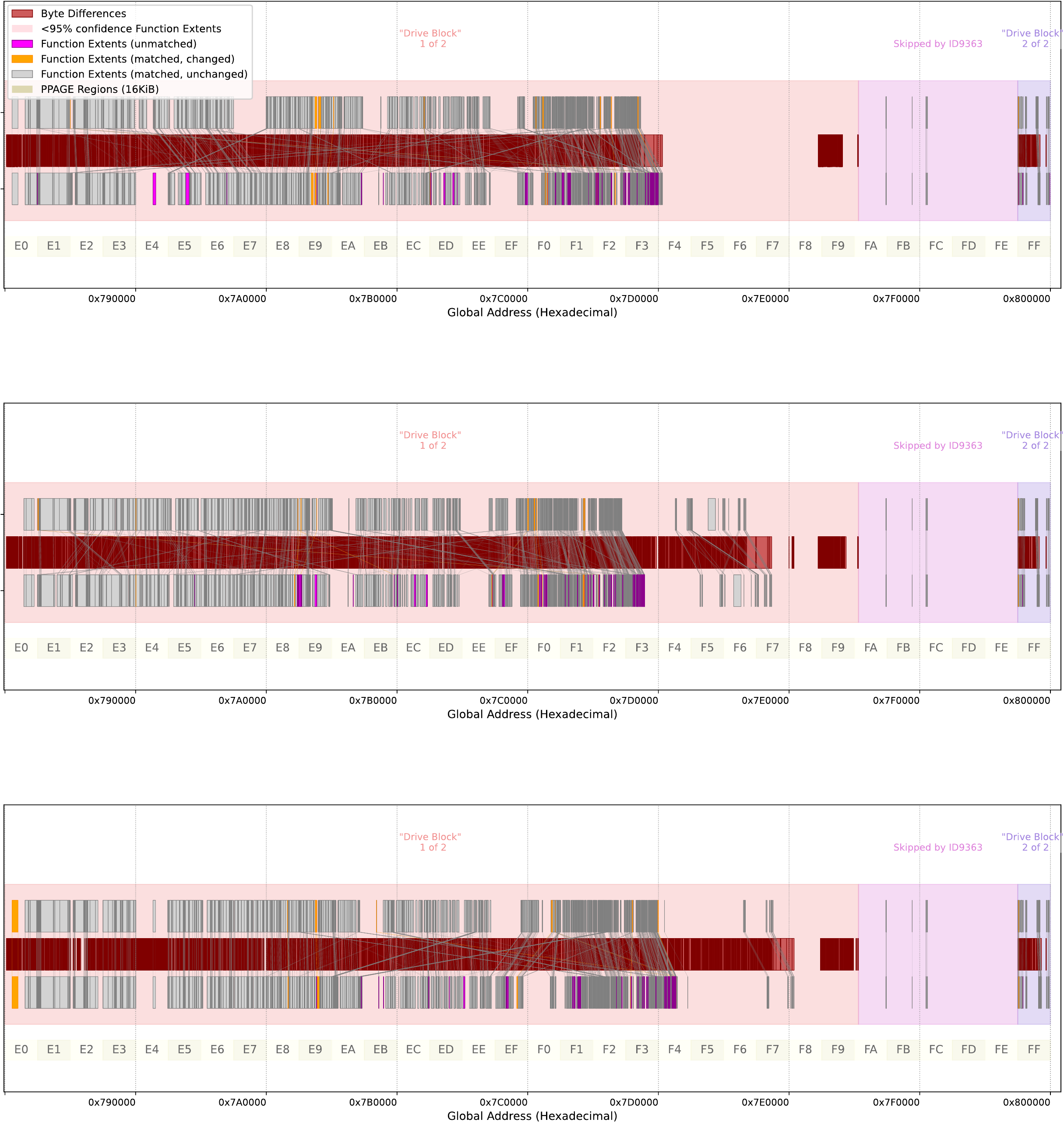}
\caption{A visualization of the QBinDiff analyses of all three of the firmware updates (when limited to the APPLICATION parts of the PFLASH), overlaid on the byte differences of each. The varying complexity of the firmwares mentioned in the introduction is clearly illustrated here (compare to Table \ref{tab:ec80targets}); target 3ec80 (top) shows the least complexity while 1ec80 (bottom) shows the most. Links between matched (by QBinDiff) functions are plotted across pre- and post-update. Despite noise from small/null function matches, the links show clear function shifting to lower addresses and numerous unmatched functions in the pre-update images (colored magenta), suggesting multiple function deletions. More conclusively: there are no (known) spurious matches and zero 'new' functions post-update: further suggesting multiple function deletions in the update.}
\label{fig:all_ecus_diff_stacked}
\end{figure*}

We ultimately created a QBinDiff \citeproc{ref-CAIDQBinDiff}{45}
analysis, the results of which are in Figure
\ref{fig:all_ecus_diff_stacked}. The key to successful QBinDiff analysis
was to `anchor' (force a match) all the interrupt handlers. This yielded
improvements over the Bindiff results. Less than 1\% of discovered
functions were low-confidence matches. Over 98\% of functions remained
unchanged, while \textasciitilde110 - \textasciitilde140 were modified
or deleted (Table \ref{tab:bindiff_results_summary}).

\begin{table}[htbp]
\centering
\footnotesize
\begin{tabularx}{\columnwidth}{@{} >{\raggedright\arraybackslash}X c c c @{}}
\toprule
  & 1ec80 & 2ec80 & 3ec80 \\
\midrule
New & 0 & 0 & 0 \\
Deleted & 104 & 127 & 123 \\
Modified & 11 & 12 & 13 \\
Unchanged & 1109 & 1100 & 1031 \\
Low Confidence & 12 & 13 & 14 \\
\bottomrule
\end{tabularx}
\caption{QBinDiff results summary of all three targets.}
\label{tab:bindiff_results_summary}
\end{table}

Three-way BinDiff analysis showed that matched functions dropped from
968 (pre-update) to 878 (post-update) across all `Z' versions,
corroborating function deletion. The patch makes very similar changes in
all three cases: of the deleted functions, 87 were common to all three
updates. The next section, \ref{changes-made-in-patch}, analyzes these
changes. Fewer functions were deleted in the 1ec80 update because it
retained J1587 Transport Protocol (TP) management functions which were
deleted in the others; presumably, because they are used due to the
J1708 interface which 1ec80 has and the other targets do not.

\begin{table}[htbp]
\centering
\footnotesize
\begin{tabularx}{\columnwidth}{@{} >{\raggedright\arraybackslash\hsize=0.8\hsize}X >{\raggedright\arraybackslash\hsize=1.4\hsize}X >{\raggedright\arraybackslash\hsize=0.8\hsize}X @{}}
\toprule
Term & Definition & BinDiff Terminology \\
\midrule
\textbf{Function Extents (matched, unchanged)} & Functions present in both binaries with identical logic and structure. & Matched, Similarity = 1.0 \\
\textbf{Function Extents (matched, changed)} & Functions present in both binaries but with some modifications to instructions or structure. & Matched, Similarity < 1.0, Confidence >= 0.95 \\
\textbf{Function Extents (unmatched)} & Functions present in one binary but not the other (new or deleted). & Unmatched (Primary/Secondary) \\
\textbf{< 95\% confidence Function Extents} & Functions that might be matched but with low confidence; ignored for high-assurance analysis. & Matched, Confidence < 0.95 \\
\bottomrule
\end{tabularx}
\caption{Table defining the categories of functions used in Figure \ref{fig:all_ecus_diff_stacked} and the analysis presented in Sections \ref{changes-made-in-patch} and \ref{exploitability-of-removed-functionality}.}
\label{tab:bindiff_terms}
\end{table}

\section{Changes Made in Patch}\label{changes-made-in-patch}

Manual BinDiff analysis, validated by call tree analysis script
(Appendix \ref{call-tree-script}), categorized all the deleted functions
in all three firmwares as one of:

\begin{enumerate}
\def\labelenumi{\arabic{enumi}.}
\item
  All J1587 PID processing present in the pre-update image (Appendix
  \ref{j1587-pid-processing}), excluding LAMP and active ABS event
  processing.
\item
  SCI2 UART EDGE interrupt handling (Appendix
  \ref{j2497-reception-architecture-of-ec80}).
\item
  Secondary J1587 features, including the diagnostic code manager
  (\texttt{sub\_E9858A}), J1587 Transport Protocol (TP) connection
  management (\texttt{sub\_EF905B}), and TP timeouts
  (\texttt{sub\_EC8116}). \footnote{J1587 TP was not functional even
    before the update, confirmed with static analysis and testing.}
\end{enumerate}

\begin{figure}[htpb]
\includegraphics[width=\columnwidth]{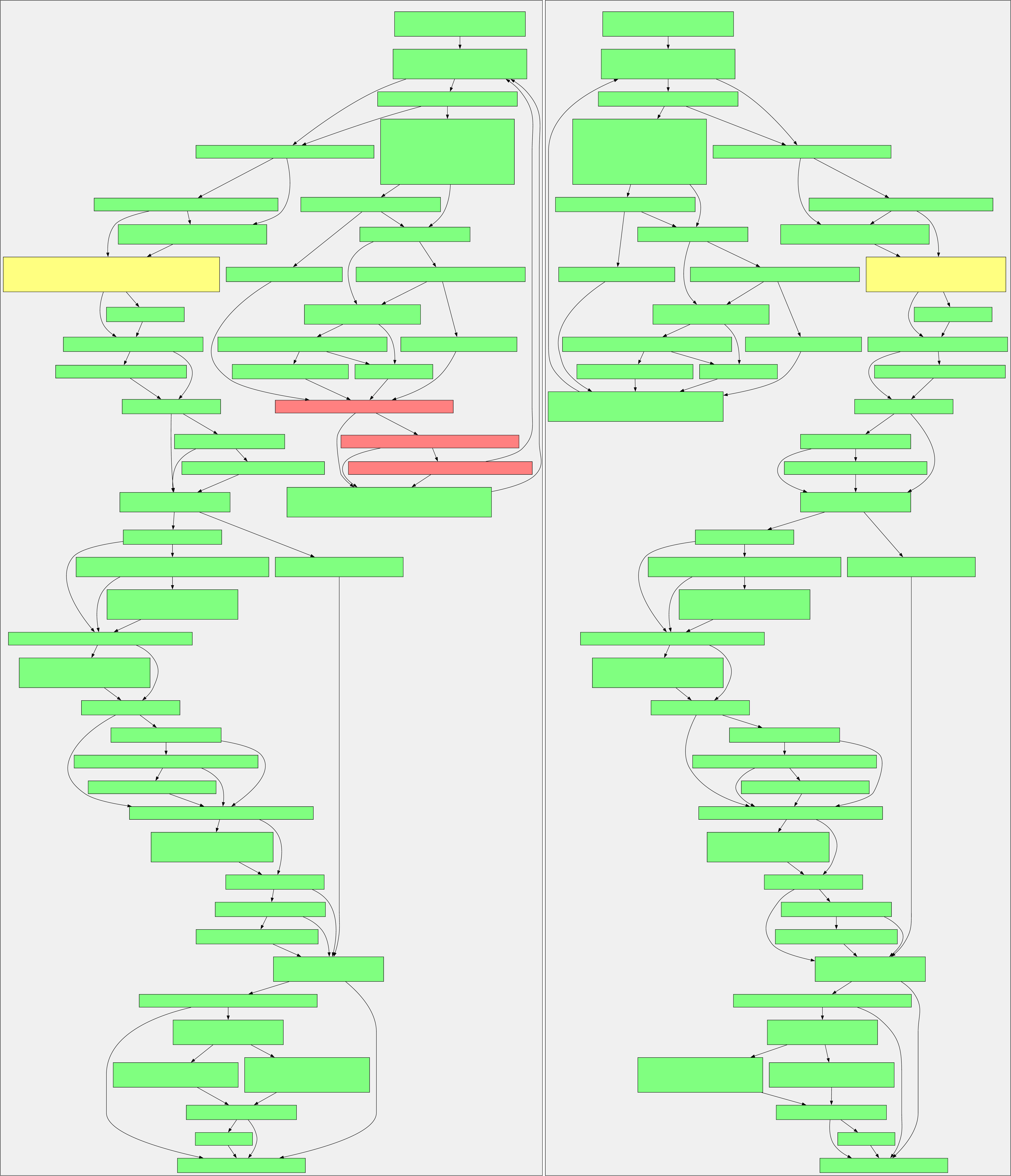}
\caption{BinDiff of the LAMP etc check function before and after update. The modified EBB has a two-line change: remove a call to \texttt{sub\_F096F} (Process PID payloads) and remove load of an argument to that call. The deleted EBBs were removed because with the call to PID processing removed there is no longer any need to handle LAMP and Active ABS MIDs at a higher priority.}
\label{fig:lampetc_check_diff}
\end{figure}

\begin{figure}[htpb]
\includegraphics[width=\columnwidth]{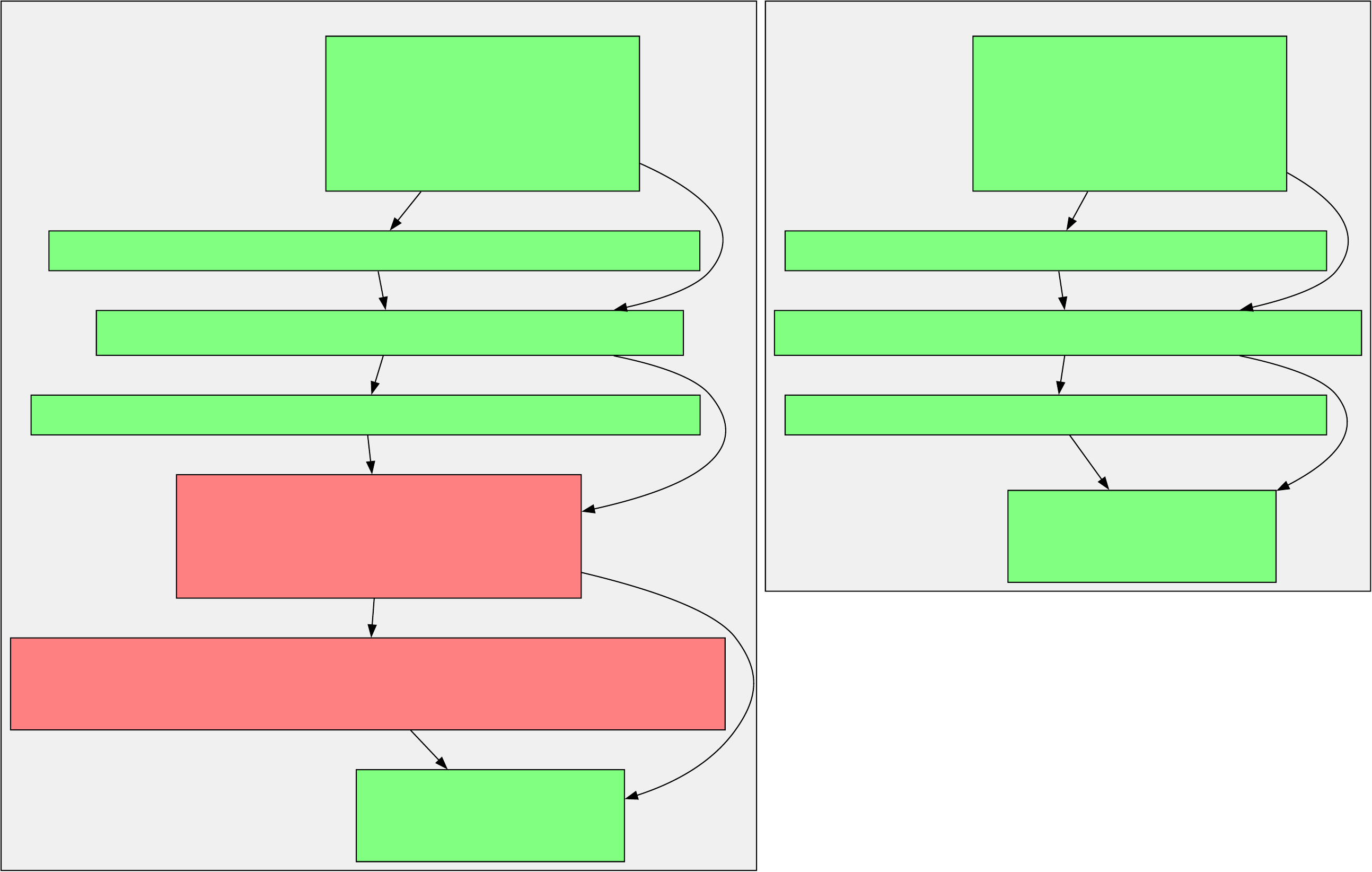}
\caption{BinDiff of the SCI2 Handler before and after update. The deleted EBBs were responsible for servicing SCI2 EDGE interrupts.}
\label{fig:sci2_handler_diff}
\end{figure}

Modified functions primarily remove calls to these deleted functions:

\begin{enumerate}
\def\labelenumi{\arabic{enumi}.}
\item
  In \texttt{sub\_EEB989} (LAMP/ABS check), the call to
  \texttt{sub\_F096F9} and subsequent PID processing is removed (Figure
  \ref{fig:lampetc_check_diff}, EBB 0xEEBA20), consequently removing all
  linked PID processing tables and functions.
\item
  In \texttt{sub\_C12B} (SCI2 ISR), the call to EDGE interrupt handling
  is removed (Figure \ref{fig:sci2_handler_diff}), effectively
  eliminating the `RXEDGIF Branch' (Figure
  \ref{fig:sci2_handler_events_buffer}) and related setup/processing.
\end{enumerate}

The bulk of the deletions were due to the removal of all PID processing
on J2497 (see Table \ref{tab:j1587_pids} for the full list). The
disassembly of all deleted functions (for target 2ec80) is provided in
Appendix \ref{appendix-disassemblies}.

\section{Exploitability of Removed
Functionality}\label{exploitability-of-removed-functionality}

We reviewed deleted functions (primarily PID handlers, Section
\ref{changes-made-in-patch}) for vulnerabilities using static and
dynamic analysis. Dynamic analysis was limited to fuzzing due to
breakpoint limitations (Section \ref{limitations-of-dynamic-analysis}).
We identified several vulnerabilities in the deleted code.

The PID 0xC2 handler was exploitable for Denial-of-Service (DoS) and
Remote Code Execution (RCE); both verified, see below. A DoS condition
could be triggered by a J2497 message like \texttt{89C2XXX}, where a
large `n' parameter (0xFE) caused a crash. For RCE, the handler
processes a length parameter specifying the total size of data following
the length parameter. The data is copied to a 0xXXXXXXXX-byte stack
buffer without length check. The return address is 0xXXXXXXXX bytes from
this buffer, creating a classic buffer overflow
\citeproc{ref-phrackbufferoverflow}{46} (see Listing of
\texttt{sub\_F1B623} in appendices) with one intervening byte: the count
of a loop executed after the copy. While J2497 frame limits restrict
direct attacker-controlled data to 18 bytes, the copy size is fully
controllable.

Furthermore, the adjacent region is attacker-controlled, as J2497 frames
are contiguous (with an injected frame-length byte) in a 0xXX-byte FIFO
buffer: an attacker can control the entire 0xXXXXX byte copy by first
sending a complete 0xXXXX byte buffer as a message (results in
additional injected length byte and added checksum byte) starting with
\texttt{89c2XXXXXXXX} which fills the buffer at 0x3BF5 (but does not get
passed to higher application layers because it is longer than 21 bytes)
then sending \texttt{89c2XXXXXXXX} by itself. The second message will
get written to the same location as the previous \texttt{89c2XXX} bytes
in the receive buffer and will be passed to higher layers (with the
remaining buffer in-tact). The wrinkle here is that any reception of
other messages on the J2497 databus will interfere with the layout in
the buffer: messages received before will move the start position of the
first and/or second messages, reducing available buffer space and
messages received in-between will move the start of the second message,
removing the adjacency of the attacker-controlled buffer. A typical
J2497 databus does contain a stream of LAMP messages and e.g.~0xC2 and
other lower priority messages. There is a means to inhibit reception:
during the development of the keyhole mitigation
\citeproc{ref-j2497keyhole}{47} the ideal standing sinusoid frequency
was discovered to inhibit message reception. An attacker can transmit
this sinusoid interference (to empty the buffer) and then transmit their
exploit in a blanking of the sinusoid. They can also abuse the Intellon
SSCP485 receiver feature where the pre-amble is superfluous to send the
two body-only messages in a sequence quicker than `normal' J2497 databus
transmitters can achieve.

The RCE was confirmed on the 2ec80 and 3ec80 targets in a bench
environment. This confirms that although the S12X has an MPU which could
inhibit execution from data buffer and/or stack areas, it was not
configured to do so. The following listing shows a simple proof of
concept with NOP placeholders. The first code block is copied to the
stack (as described above); the return address is overwritten with the
address of the first code block in the receive FIFO 0x3BF5. Execution
flows into the receive buffer original copy of the code block and jumps
over the return address and intervening byte there to use an additional
area for more code in the buffer.

\begin{Shaded}
\begin{Highlighting}[]
\CommentTok{\# REDACTED FOR PREPRINT}
\end{Highlighting}
\end{Shaded}

This PoC uses the receive FIFO buffer at 0x3BF5 as the location for
hosting the executable code and only queues one message (the
\texttt{{[}0:20{]}} slice above). If other bytes are received before the
\texttt{SEI} is executed they will be written into the buffer after
0x3BF5+20. The interfering sinusoid mentioned above can be reasonably
effective at accomplishing this. A more robust approach could use the
stack only (if the SP address in the 0xC2 handler were known
\footnote{it was not for us. Because there was no dynamic analysis
  possible this PoC was developed iteratively in a crash/no-crash loop})
or an approach could send multiple messages and skip over the injected
length bytes. Long-running payloads will require feeding the watchdog
peripheral and there are many examples of this in the firmware.

The PoC leaves 66 bytes of usable payload; this is enough to transmit a
valid J1939 \citeproc{ref-saej1939}{36} CAN frame. The basic form of
this is 85 bytes long using MOVB for data move. This can be reduced to
62 bytes by using STD 2,X+ instead. And it can be golfed into a 43 byte
payload by abusing the stack for PULX; however, using PULX is not
practical on the targets which use the XGATE coprocessor during their
runtime (such as 2ec80) because the stack is shared. The payload can
also be reduced to 39 bytes by re-using the firmware's CAN-sending
function \texttt{sub\_E9919A}. This was confirmed using the HSW12
assembler \citeproc{ref-hsw12}{41}. Also possible is misconfiguring the
CAN transceiver to a bad bps which will crash that CAN bus segment and
this a much smaller payload. This will silence all ECUs on the bus --
until bus recovery fixes the problem, if at all. More is possible. As is
always the case with RCE: anything is possible
\citeproc{ref-masterjun_smw_rce}{48}. The space limitation here is also
only a minor problem since an attacker can find many RAM areas to write
longer payloads `in pieces' as well.

Returning to the vulnerabilities in the other deleted code, the PID 0xED
\footnote{which is Trailer VIN PID, not the previously allocated PID:
  Entry Assist Control \#1.} handler can also crash the ECU. Sending a
rapid sequence of messages with a large length parameter triggered this
vulnerability. Bench and closed-track testing confirmed that greater
than 10 repeated \texttt{89EDXX} messages triggered the crash (128x was
used to be sure). The PID 0xED handler (\texttt{sub\_F2BA46}) causes
this by copying the PID payload (length-specified) into a static buffer
at 0x5BDC without bounds checking (see Listing for
\texttt{sub\_F2BA46}s). This buffer has a XXX-byte safe limit before
overwriting critical structures at 0x5BF6 and 0x5BF8. These locations
are used by \texttt{sub\_F38241} for a conditional \texttt{memcpy}
(\texttt{sub\_E642}), with the condition seemingly controllable via the
PID 0xB4 handler. Like the 0xC2 handler, the attacker controls copy size
and can control the adjacent bytes in the receive FIFO buffer.

Theoretically, sequential 0xED and 0xB4 PID payloads (potentially in one
J2497 message) could establish a write primitive. The write primitive
pointer and data are XXX and XX bytes from the 0xED handler's buffer
start. Successful exploitation requires careful J2497 receive buffer
grooming to overcome the XX-byte control limit. This write primitive
makes RCE feasible, given unused RAM regions and fixed-address function
pointers in the firmware \footnote{It would also be possible to send CAN
  data with a write primitive and without RCE.}.

Deleted SCI2 EDGE interrupt code theoretically presented a write-where
primitive. Exploitation involves: 1) Triggering an EDGE interrupt and
timeout to partially reset the receive buffer (by \texttt{sub\_F1B035}),
setting a buffer full flag (see Table \ref{tab:control_bits}) but
leaving the write offset (0x3C4B) un-reset. 2) Consuming a subsequent
message clears the buffer full flag without resetting the offset,
potentially exceeding the 0xXXXXXXXX buffer limit. 3)
\texttt{sub\_EC803E} (append data) contained a write-before-check flaw.
Data was written to the receive buffer (0x3BF5) at the current offset
(0x3C4B) before boundary validation. This allowed writes beyond the
limit. 4) Positioning the write offset at +0xXX and sending a X-byte
frame allowed setting arbitrary upper bound (0x3C49) and current offset
(0x3C4B) values. Subsequent data would be written to the new offset
address.

The PID 0xC7 handler contained a hardcoded password check. Receiving the
matching password disables traction control, as per J1587 specification
\citeproc{ref-saej1587}{22}. Bench testing confirmed that message
\texttt{ACC7XXXXX} satisfies this password check \footnote{REDACTED FOR
  PREPRINT}.

\begin{table}[htbp]
\centering
\footnotesize
\begin{tabularx}{\columnwidth}{@{} l >{\raggedright\arraybackslash}X >{\raggedright\arraybackslash}X @{}}
\toprule
\textbf{Component} & \textbf{Vulnerability} & \textbf{Impact} \\
\midrule
PID 0xC2 & Buffer Overflow & DoS (Verified), RCE (Verified) \\
PID 0xED & Unbounded Copy & DoS (Verified), RCE (Theoretical) \\
PID 0xC7 & Hardcoded Credential & Auth Bypass (Verified) \\
SCI2 EDGE & OOB Write & DoS (Theoretical), RCE (Theoretical) \\
\bottomrule
\end{tabularx}
\caption{Summary of vulnerabilities in removed functionality.}
\label{tab:vuln_summary}
\end{table}

Exploits of these RCE vulnerabilities (verified and theoretical) will
depend on the specific `Z' firmware version memory layout. The ID9363
updater contains 13 `Z' versions (covering a total
\textasciitilde450,000 units \citeproc{ref-nhtsa_recalls_datahub}{12});
however, exploits of the DoS vulnerabilities and hardcoded credential
will not. Testing shows they worked in all three tested `Z' versions and
likely affect all versions.

Reverse engineering is not required to discover all these
vulnerabilities. A simple fuzzing script discovered the PID 0xC2 and
0xED crashes. These were verified on the bench and in-motion (Section
\ref{notes-on-in-motion-vehicle-tests}). See Appendix
\ref{fuzzing-script} for the script used.

\section{Notes on In-Motion Vehicle
Tests}\label{notes-on-in-motion-vehicle-tests}

\begin{figure}[htpb]
\includegraphics[width=\columnwidth]{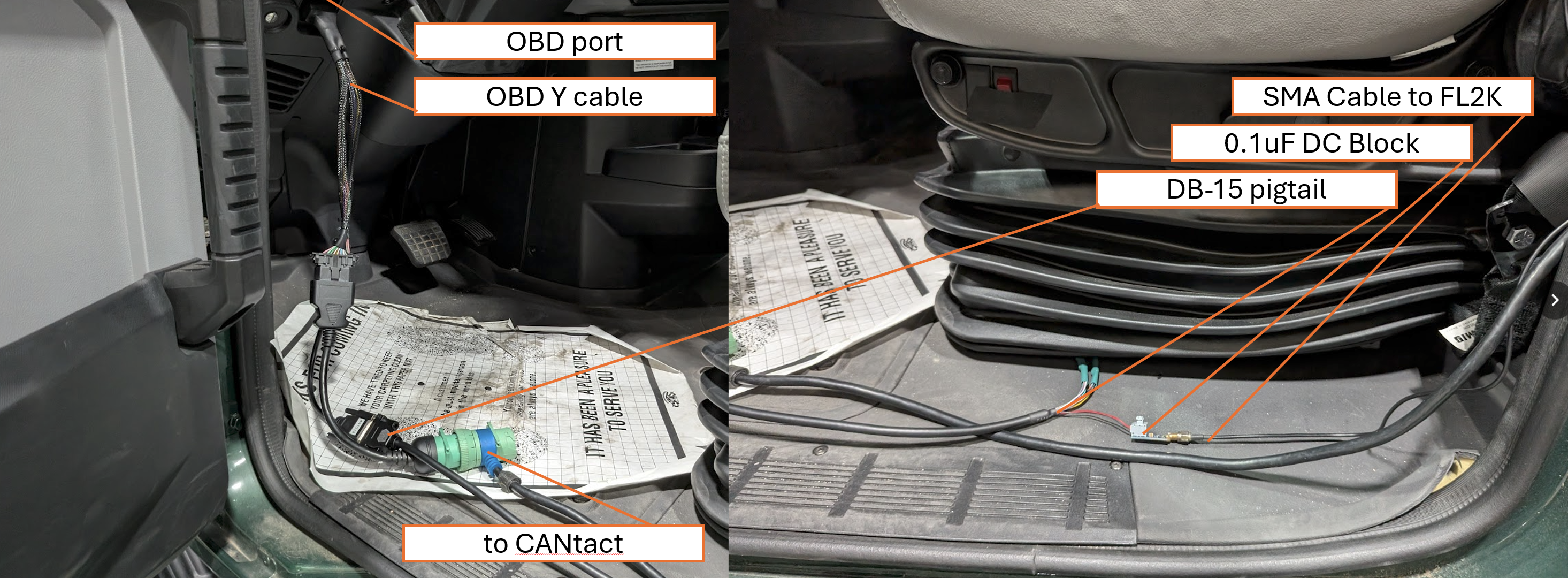}
\caption{Right: a 10uF DC block resulting in AC-coupling of the FL2K SDR (not pictured) connected to a power amplifier (not pictured). Left: Connection of the same to the in-cab on board diagnostic (OBD) port.}
\label{obdanddcblock}
\end{figure}

\begin{figure}[htpb]
\includegraphics[width=\columnwidth]{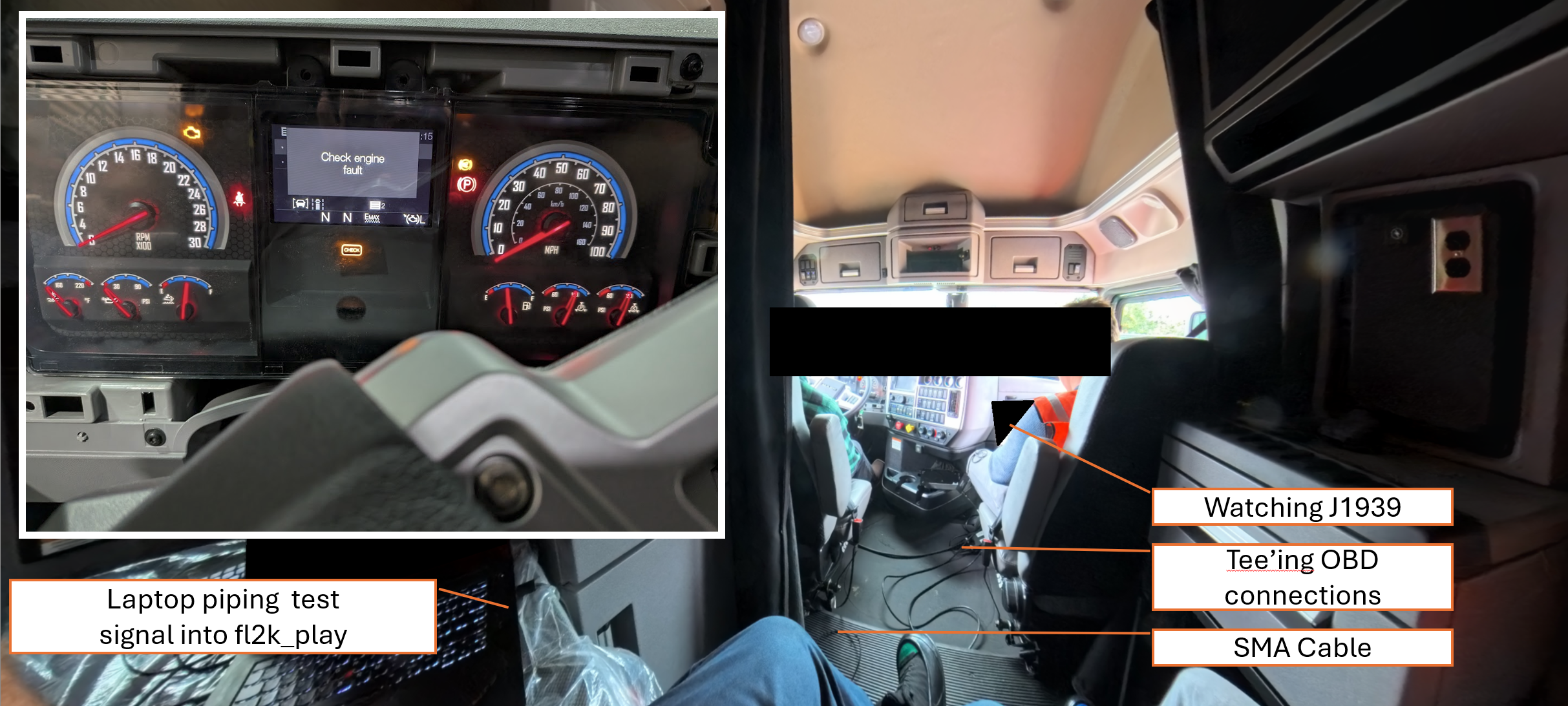}
\caption{Top Left: failure state of instrument cluster after any and all the ECU crashes described in Section \ref{notes-on-in-motion-vehicle-tests} while still in-motion. Wide: in-cab picture of in-motion vehicle test with black bar redacting.}
\label{speedoandcabredacted}
\end{figure}

Onsite testing used a host-provided tractor that did not perfectly match
the source vehicle of the vulnerable firmware but whose EC80 part number
matched perfectly -- because the host did not have any EC80s that were
not already updated by ID9363. We manually flashed the target ECU with
the vulnerable firmware via BDM. Then the target ECU was part-calibrated
to the tractor using OEM tools, and baseline ABS functionality was
confirmed via hard braking before further testing.

Testing focused on the crashes in the handlers for J1587 PIDs 0xC2 and
0xED, identified during bench testing. We injected J2497 signals using
an FL2K Software Defined Radio (SDR) transmitter
\citeproc{ref-tedyapofl2k}{49} \citeproc{ref-osmofl2k}{50} and power
amplifier. Due to insufficient J2497 filtering at the diagnostic port,
we were able to use a simple in-cab setup: AC-coupled via a 10uF
capacitor to the diagnostic port's VBAT pin \footnote{this was confirmed
  by observing Trailer ABS fault dash light status and 0xF001 PGN with
  \texttt{python\ -m\ can.viewer\ -i\ cantact\ -c\ 0\ -b\ 500000\ -\/-filter=00F00100:00FFFF00}
  (using python-can \citeproc{ref-pythoncan}{51} and CANtact
  \citeproc{ref-cantact}{52}).} (Figure \ref{obdanddcblock}). As shown
in \citeproc{ref-gardiner2022disclosure}{4}, this simulates wireless
injection \citeproc{ref-CVE-2022-26131}{6}.

We conducted in-motion tests below 5 mph and at \textasciitilde9 mph ---
fast enough to trigger ABS pulsing but safe for occupants. At 9 mph,
after observing the ECU crash via CAN, the driver initiated a hard brake
to check for ABS pulsing. In all cases, CAN traffic ceased. ECU recovery
always required a battery disconnect \footnote{This is a relevant
  condition because some tractors are configured without a battery
  disconnect switch, requiring a manual removal of the battery leads or
  a manual cycling of a fuse to recover the ECU.}. The crash
consistently caused loss of speedometer, steering assist, and shifting
(Figure \ref{speedoandcabredacted}). The results are summarized in Table
\ref{tab:in-motion-tests}. The scripts used to trigger the 0xED and 0xC2
DoS vulnerabilities can be found in Appendices
\ref{ed-pid-handler-crash}, and \ref{c2-pid-handler-crash},
respectively.

One difference emerged: in the PID 0xED case, the EC80 stopped streaming
signal data but still responded to UDS TP requests. This was not
observed in the PID 0xC2 case or during PID 0xED bench testing.

We repeated these in-motion tests on a second vehicle: owned by a fleet
and from a different OEM than the previous. This EC80 was `factory
installed' and was not yet updated by `ID9363'. During these tests we
were able to test also the 0xC7 `backdoor' impacts. In this vehicle the
impact of crashing the ABS controller did not include loss of
speedometer nor steering assist nor shifting; but other impacts were the
same. This was also true for the 0xC7 `backdoor'.

\begin{table}[htbp]
\centering
\footnotesize
\setlength{\tabcolsep}{1.25pt}
\renewcommand{\arraystretch}{1.15}

\begin{tabular}{
@{}
p{0.75cm}  %
p{1.0cm}  %
p{1.0cm}  %
p{0.75cm}  %
p{3.5cm}  %
p{0.75cm}  %
@{}
}
\toprule
\textbf{Vuln.} &
\textbf{Target} &
\textbf{Speed} &
\textbf{UDS} &
\textbf{Notes} &
\textbf{ABS Disabled} \\
\midrule

0xED & 1ec80 & $<$ 5 mph & Yes &
Loss of shift, steering assist, speedo. Multiple cluster faults &
 -- \\
0xED & 3ec80 & $<$ 5 mph & No &
Mulitple Cluster faults. &
 -- \\

0xED & 1ec80 & $>$ 5 mph & Yes &
Loss of shift, steering assist, speedo &
\textbf{Yes} \\
0xED & 3ec80 & $>$ 5 mph & No &
Mulitple Cluster faults. &
\textbf{Yes} \\

\midrule

0xC2 & 1ec80 & $<$ 5 mph & No &
Loss of speedo, steering assist &
-- \\
0xC2 & 3ec80 & $<$ 5 mph & No &
Mulitple Cluster faults. &
-- \\

0xC2 & 1ec80 & $>$ 5 mph & No &
Loss of shift, steering assist, speedo &
\textbf{Yes} \\
0xC2 & 3ec80 & $>$ 5 mph & No &
Mulitple Cluster faults. &
\textbf{Yes} \\

\midrule

0xC7 & 3ec80 & $>$ 5 mph &
No &
Multiple Cluster faults including ``Dyno Mode''. Loss of shifting in Reverse &
\textbf{Yes} \\

\bottomrule
\end{tabular}
\caption{In-motion test results (In all cases streaming data stopped).}
\label{tab:in-motion-tests}
\end{table}

\section{Random Noise Sufficiency}\label{random-noise-sufficiency}

The EDGE SCI2 interrupt vulnerability could obviously have been
triggered by noise: constant framing errors and/or parity errors could
cause false 0x00 bytes to fill the receive buffer and edge interrupts
firing coincident with these could cause the partial reset and eventual
overwrite of pointers. It is not as clear for the PID processing
handlers' vulnerabilities.

As was mentioned in Appendix \ref{j2497-reception-architecture-of-ec80},
the only receiver error checking is if the checksum is consistent. These
checksums are the two's complement of the sum of all bits in the
message. The checksum is not able to detect many-bit errors, and
e.g.~fails to detect `symmetric' errors (pairs of bit flips). It is easy
to imagine the checksum failing to prevent reception of invalid messages
in noisy conditions.

There are not any established models of bit errors or of spurious data
reception of the Intellon SSCP485 \citeproc{ref-sscp485}{24} that we are
aware of. However, from previous research
\citeproc{ref-Gardiner2022Mitigating}{53} it is known to be possible for
a message on the bus to have several of its bits flipped and/or be
truncated on reception in the presence of noise.

It is very typical of trailer brake controllers to stream PID 0xC2 and
so it could be imagined that bit flips of an existing 0xC2 message could
cause a large value of `n' and hence the same DoS crash we observed in
Section \ref{notes-on-in-motion-vehicle-tests}.

Trailer brake controllers do not typically stream 0xED messages
\footnote{although they can send them on power up and on request at
  power-up of the tractor}. Nor do they stream 0xC7 messages. It is,
however, possible that the 0xC2 (or other streaming messages) get bit
flipped to 0xED or 0xC7 and truncated to the right length. This could
not be reasonably described as a likely possibility though.

\section{Related Work}\label{related-work}

Binary analysis of bare-metal firmware often faces unique constraints.
While \citeproc{ref-tsang2024ffxe}{54} uses emulation, we employ
minimum-viable concolic analysis where emulation was unavailable. We
address BinDiff's small-function mismatch issues
\citeproc{ref-marcelli2022how}{55} by applying QBinDiff
\citeproc{ref-mengin2020qbindiff}{56} with manual interrupt handler
anchoring to enable useful results. Furthermore, our findings confirm
that silent security patching \citeproc{ref-ruge2020frankenstein}{57} is
occuring also in heavy vehicle firmware.

Seminal work \citeproc{ref-koscher2010experimental}{58} identified
``standard'' vulnerabilities like buffer overflows that, we show,
persist in modern components. We extend J1708 security research
\citeproc{ref-burakova2016truck}{59} to J2497 and apply the wireless
write attack \citeproc{ref-Gardiner2022Mitigating}{53} to tractor
equipment. Complementing J1939 diagnostic research
\citeproc{ref-chatterjee2024exploiting}{60}, we demonstrate that
mission-time (as distinct from diagnostics) data processing defects in
J2497 (as opposed to J1939) directly impact tractor safety.

\section{Generalizable Contributions}\label{generalizable-contributions}

Our work provides several contributions that generalize across different
architectures and target devices. First, we demonstrate that hiveplots
can be effectively used to survey the current disassembly of a binary.
This visualization technique is particularly useful for identifying
`configuration' issues and spotting segmented areas in bare-metal
firmware images. Second, we found that the use of manual anchoring in
QBinDiff for each of the interrupt handlers is a critical step in
achieving useful and accurate binary diff results in its application to
bare metal firmware. Finally, our approach of performing live RAM
dumping is generally applicable to any case where a debugger is not
fully functional. This technique is critical for discovering dynamically
resolved functions, particularly when function pointers or their offsets
are stored exclusively in RAM.

\section{Conclusions}\label{conclusions}

This work demonstrates binary differential analysis on legacy automotive
architectures, using heuristics and symbolic analysis to overcome
tooling limitations. Our analysis reveals several vulnerabilities were
patched by the ID9363 update.

While the vendor emphasized safety compliance regarding random noise,
our analysis confirms a broader security impact. Random line noise is
unlikely to generate the valid J1587 frames and frame sequences required
to trigger these vulnerabilities. Specifically, the patch removes the
tractor ECU's J1587 parsing stack (except LAMP/ABS event handling),
eliminating vulnerable handlers for PIDs 0xC2, 0xED, and 0xC7.

We classify the removed functionality as critically flawed, containing
unauthenticated memory corruption (PIDs 0xC2, 0xED) and hardcoded weak
authentication (PID 0xC7). In-motion testing validated the severity:
exploiting the PID 0xC2 buffer overflow caused immediate ABS Denial of
Service (DoS), physically manifesting as loss of steering assist,
speedometer, shifting and ABS pulsing. The PID 0xED handler induced the
same DoS via memory corruption and the 0xC7 handler also via abuse of
weak authenticated functionality. Bench testing confirmed RCE via the
memory corruption in the PID 0xC2 handler.

The ID9363 update mitigates risks from attackers with wireless adjacency
or other forms of J2497 network access, improving the compliance to
requirements NGW-S-001 through -005
\citeproc{ref-gardiner2024security}{29}. Bendix's proactive recall and
firmware update represent a commendable and responsible action. This is
particularly important as users are believed to prioritize safety
recalls over security patches. However, the lack of distinct CVE
assignments for these flaws may obscure the patch's security
criticality. What is recommended -- by ISO/IEC 29147
\citeproc{ref-iso29147}{61} and CISA \citeproc{ref-cisa2023shifting}{62}
-- is that vulnerabilities which are patched should be communicated to
users so that they can make their own informed risk calculations.

The features removed align with the security architecture for tractor
J2497 reception \footnote{for tractor J2497 transmission the same
  sources recommended that the tractor mitigate attacks on older trailer
  equipment; this patch does not appear to add any such mitigations.}
detailed in all of:

\begin{enumerate}
\def\labelenumi{\arabic{enumi}.}
\item
  the ATA TMC position paper on next generation tractor trailer
  interfaces \citeproc{ref-tmc2024nextgen}{63}: ``Pertaining to PLC
  communications as described in SAE J2497: only the MID 10 and 11 lamp
  messages, MID 125 J2497 identification, and MID 87 active ABS event
  shall be permitted on new tractor and trailer equipment\ldots{}''
\item
  the CISA mitigations in advisory ICSA-25-021-03
  \citeproc{ref-cisa2025advisory}{64}: ``To most effectively mitigate
  general vulnerabilities of the powerline communication, any
  {[}tractors{]} utilizing J2497 technology should disable all features
  where possible, except for backwards-compatibility with LAMP ON
  detection only.''
\item
  the latest version of SAE J2497 \citeproc{ref-sae2026j2497}{1}: ``For
  tractor PLC units, it is recommended that the software implements only
  reception of the LAMP ON and OFF messages \ldots{} All other reception
  of messages should be performed on other network types.''
\end{enumerate}

Given the potential for a) malicious J2497 signal injection and b)
compromised J2497-capable telematics units, we strongly advocate that
all tractor J2497-receiving equipment, especially brake controllers,
adopt the security-conscious approach demonstrated by the EC80 update.

\section*{Acknowledgments}\label{acknowledgments}
\addcontentsline{toc}{section}{Acknowledgments}

We would like to thank our colleagues at the National Motor Freight
Traffic Association Inc.~whose expertise and dedication have been
invaluable to the completion of this research. Particularly Anne Zachos
for the continuous assistance in analysis and onsite testing.

We would like to thank AIS for access to their Class 8 vehicle multiple
times during this research. Thank you to Hannah Silva and Jesse Norton
for sharing S12X development materials and EC80 experience. Thank you to
Chris York for the trailhead. Thank you to Jonatan Mars for critical
support at a critical time. We would also like to thank many industry
engineers for their support of this research -- the rest of whom would
prefer not to be named. Without the support of the engineers at these
vehicle OEMs and suppliers this work would not be possible. Last but not
least, this work would also not be possible without the support of the
member fleets of the NMFTA Inc.~

We used various large language models (Gemini \citeproc{ref-gemini}{65}
2, 2.5-pro, 3-pro) for tasks such as: drafting Python code and Mermaid
diagrams. We acknowledge the open-source tools used: Pandoc
\citeproc{ref-pandoc}{66}, Matplotlib \citeproc{ref-matplotlib}{67},
Hiveplotlib \citeproc{ref-hiveplotlib}{68}, Zynamics BinDiff
\citeproc{ref-bindiff}{43}, Quarkslab QBinDiff
\citeproc{ref-CAIDQBinDiff}{45}, Quokka \citeproc{ref-quokka}{69},
Python 3 \citeproc{ref-python}{70}, python-can
\citeproc{ref-pythoncan}{51}, py-hv-networks
\citeproc{ref-pyhvnetworks}{71}, RP1210 python
\citeproc{ref-rp1210_dfieschko}{72}, Osmocom FL2K
\citeproc{ref-osmofl2k}{50}, Linkerscope \citeproc{ref-linkerscope}{73},
Mermaid \citeproc{ref-mermaid}{74}, and Graphviz
\citeproc{ref-graphviz}{75}.

\section*{Open Science -
Availability}\label{open-science---availability}
\addcontentsline{toc}{section}{Open Science - Availability}

The complete code listings of scripts mentioned in the paper are
available in the Artifact Appendix of this paper
https://doi.org/10.5281/zenodo.19664811. With the exception of the
QBinDiff analysis which will be made available later as `howto' document
pull request to the Quarkslab QBinDiff repository docs.

None of the firmware images used in the analysis will be made available.
The reverse engineering of this work focused only on the machine code
patched by the firmware update. This work makes no claims on the
presence nor absence of vulnerabilities in the remainder of the firmware
domains; hence, there is insufficient guarantee of safety of sharing the
firmwares. In lieu of the firmware, we provide the following in the
Artifact Appendix:

\begin{enumerate}
\def\labelenumi{\arabic{enumi}.}
\tightlist
\item
  disassembly listings (with instruction bytes included) for the deleted
  and select modified functions only
\item
  PID Handler and Context Tables deleted in the update
\item
  an .s19 file containing only the above plus the interrupt vectors for
  the application
\end{enumerate}

The ID9363 updater executable will not be made available because it
contains copies of all the new firmware images.

The ID9363 firmware update CAN logs will not be made available because
they contain the entire updated firmware in the clear as well as
seed-key exchange pairs which cannot be guaranteed to be insufficient to
reconstruct the seed-key routines in closed-form.

\section*{Ethical Considerations}\label{ethical-considerations}
\addcontentsline{toc}{section}{Ethical Considerations}

This research prioritizes disclosure and minimizes potential harm. Prior
to commencing the technical analysis, in early 2025, the manufacturer of
the brake ECU was contacted and informed of our intent to conduct
research on their firmware update for the safety recall and requested to
collaborate. Again, as the analysis neared completion, the manufacturer
was contacted in an attempt to provide them with an overview of our
findings. Two of the three OEMs have received the results in this paper.
NHTSA and Transport Canada have also been informed of this research and
results.

Crucially, this paper discusses only functions and vulnerabilities
addressed by the manufacturer's safety recall. Only deleted function
code is shared. Limited side-by-side code modifications are shared to
limit disclosure of code in the field. The PoC provided can be used to
confirm RCE but does not disclose the details of CAN bus or other
physical control by the ECU. The updated firmware has been deployed on
\textasciitilde310,000 units \citeproc{ref-nhtsa_recalls_datahub}{12} at
the time of this writing. These vulnerabilities are confirmed absent in
updated firmware versions Z300822, Z302578, Z302579 (one for each
affected OEM). Thus, disclosing these vulnerabilities introduces no new
risks. By focusing on already-patched issues, this research aims to
contribute to the broader understanding of automotive security without
inadvertently exposing users to unmitigated risks.

Finally, the in-vehicle testing was conducted in a controlled
environment, utilizing a closed test track for in-motion vehicle
testing. All testing was performed with the explicit cooperation and
support of an OEM, ensuring that safety protocols were rigorously
followed and that no public roads or operational vehicles were subjected
to unverified or potentially hazardous conditions.

\section*{References}\label{references}
\addcontentsline{toc}{section}{References}

\protect\phantomsection\label{refs}

\FloatBarrier

\appendix
\counterwithin{figure}{section}
\counterwithin{table}{section}
\clearpage

\section{Appendix: ID9363 Update Process Further
Details}\label{id9363-update-process-further-details}

The update process is executed by the ID9363 updater executable, where
we captured the J1939 \citeproc{ref-saej1939}{36} traffic sent between
it and the target ECU. The update process uses UDS
\citeproc{ref-iso14229}{8} on J1939 \citeproc{ref-saej1939}{36}; in all
cases we observed the traffic was unicast, default priority, between the
vehicle diagnostic adapter at address 0xF9 and the brake controller at
0x0B: resulting in the two extended (29-bit) CAN IDs 0x18DA0BF9 and
0x18DAF90B. The CAN traffic was filtered on those two IDs and processed
with Wireshark \citeproc{ref-wireshark}{76} to reconstruct the ISO-TP
multi frame messages used by UDS and then these were analyzed to
reconstruct the ID9363 update process, a simplified summary of which is
reproduced in Table \ref{tab:update_steps}:

\begin{table}[htbp]
\footnotesize
\rowcolors{2}{gray!10}{white}
\begin{tabular}{@{}lp{0.55\columnwidth}@{}}
\toprule
1. Tester Present & 2x Tester Present requests; many more periodically and all omitted below. \\
2. Read DIDs (Start) & Read Data By Identifier requests for: 0xF18A, 0xF192, 0xF18C, 0xF194, 0xFDEA, 0xFDA0, 0xFDE8. \\
3. Auth to Programming & Diagnostic Session Control (0x02) followed by Security Access (0x05) seed-key exchange. \\
4. Fingerprint & Write Data By Identifier (0xF184) to record Tool ID (contains "ID9363"). Always follows auth step; omitted in the following. \\
5. Download Block 1 of 2 & RoutineControl (Erase), RequestDownload, TransferData. Writes 0x68800 bytes to PFLASH (0x780000 - 0x7E8800). \\
6. Download Block 2 of 2 & RoutineControl (Erase), RequestDownload, TransferData. Writes 0x3800 bytes to PFLASH (0x7FC000 - 0x7FF7FF). \\
7. Checksum \& Reset & RoutineControl (0xFF01 checkProgrammingDependencies) without any extra parameters, followed by Hard Reset. \\
8. Auth to Programming & Re-enter Programming Session (0x02) followed by Security Access (0x05). And fingerprint. \\
9. Download Dataset & RoutineControl (Erase), RequestDownload, TransferData. Writes 0xC00 bytes of dataset/calibration data to EEE buffer at 0x13F400. \\
10. Checksum \& Reset & RoutineControl (0xFF01) with extra parameters, followed by Hard Reset. \\
11. Auth to Extended & Enter Extended Session (0x03) followed by Security Access (0x01). And fingerprint. \\
12. Write Proprietary DID & Write Data By Identifier (0xFDA0) with values from Step 2. \\
13. Write Customer PN & Write Data By Identifier (0xF191 vehicleManufacturerECUHardwareNumberDataIdentifier). \\
14. Auth to Extended & Enter Extended Session (0x03) followed by Security Access (0x07). And fingerprint. \\
15. Write Proprietary 'Revision' DID & Write Data By Identifier (0xFDEA) with revision data (e.g., 'R011'). \\
16. Update Proprietary DID & Read and Write Data By Identifier (0xFDE8). \\
17. Auth to Extended & Enter Extended Session (0x03) followed by Security Access (0x03). And fingerprint. \\
18. 'Commit Step' \& Reset & Proprietary Requests: 0xFE5C, 0xFE5D, 0xFE6A followed by hard reset. \\
19. Clear DTCs & \\
20. Read DIDs (End) & Read Data By Identifier requests (0xF192, 0xF194, 0xFDEA, 0xF18C) to verify update. \\
\bottomrule
\end{tabular}
\caption{ID9363 Update Process Steps.}
\label{tab:update_steps}
\end{table}

\FloatBarrier

\section{Appendix: J2497 Reception Architecture of
EC80}\label{j2497-reception-architecture-of-ec80}

\begin{figure*}[htpb]
\centering
\includegraphics[width=\textwidth]{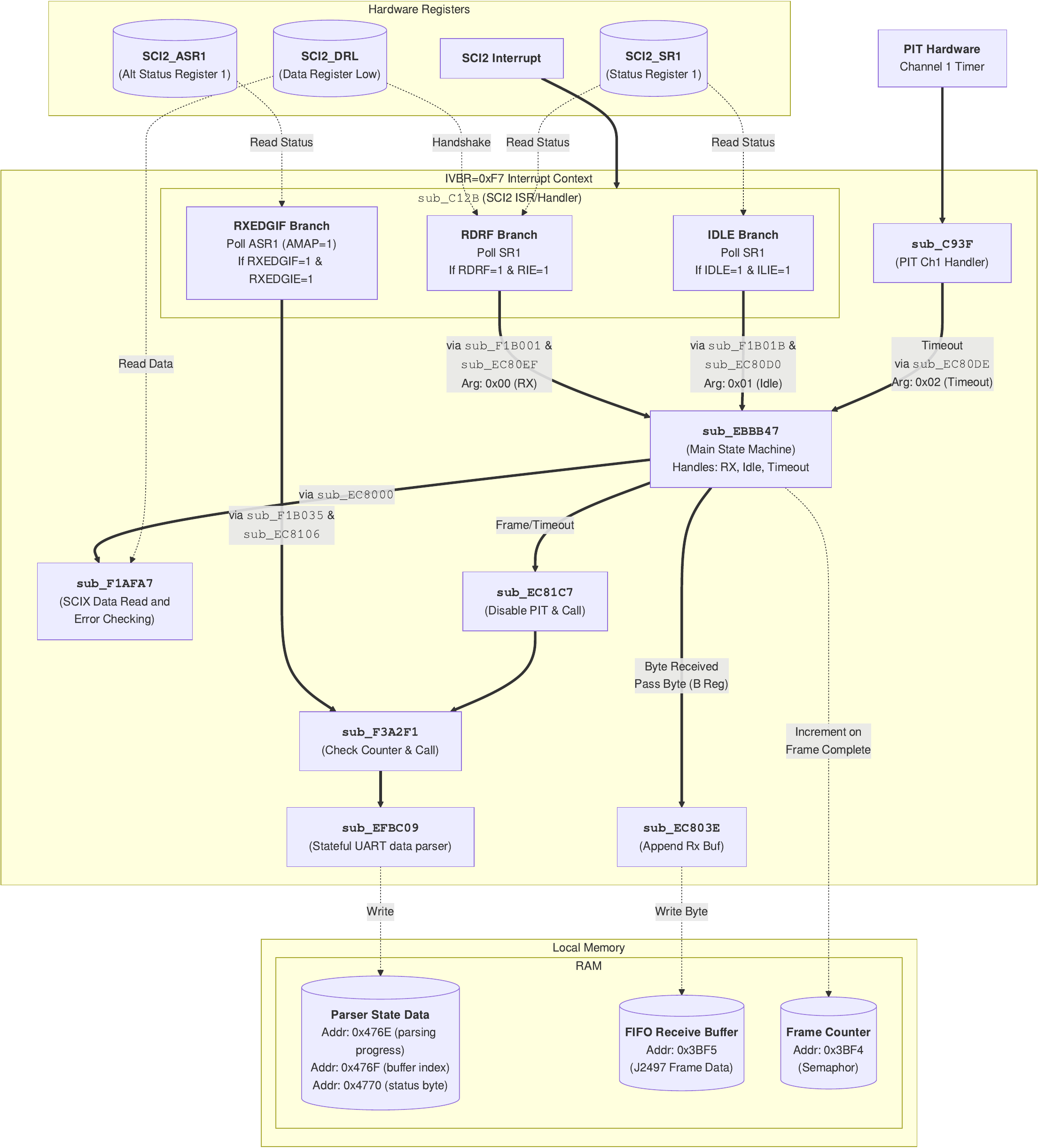}
\caption{Dataflow diagram visualization of the handling of J2497 UART peripheral events: edge, idle, data and also of PIT peripheral events used for timeouts.}
\label{fig:sci2_handler_events_buffer}
\end{figure*}

The reception is driven by events from the SCI2 and PIT hardware
peripherals as illustrated in Figure
\ref{fig:sci2_handler_events_buffer}, PIT being used to manage timeouts.
SCI2 is a UART that supports several interrupt event types and the
receiver state machine makes use of `data ready' (RDRF), IDLE, and edge
(RXEDGIF) events. The entire state machine is distributed over several
functions which all access address 0x3C52, a set of control bitfields,
summarized in Table \ref{tab:control_bits}.

\begin{table}[htbp]
\centering
\footnotesize
\begin{tabular}{@{} l l p{0.5\columnwidth} @{}}
\toprule
Bit Mask & Control Name & Function Address(es) \\
\midrule
\texttt{0x01} & \textbf{PENDING} & \texttt{sub\_EBBE13} \\
\texttt{0x02} & \textbf{IGNORE\_RX} & \texttt{sub\_EBBB47} \\
\texttt{0x04} & \textbf{TIMEOUT} & \texttt{sub\_EBBB47}, \texttt{sub\_EBBE13}, \texttt{sub\_EC80B5}, \texttt{sub\_EBBEDB} \\
\texttt{0x08} & \textbf{BUF\_DONE} & \texttt{sub\_EBBB47}, \texttt{sub\_EBBE13}, \texttt{sub\_EC80B5} \\
\texttt{0x10} & \textbf{BUF\_FULL} & \texttt{sub\_EBBB47}, \texttt{sub\_EC803E} \\
\texttt{0x20} & \textbf{EDGE\_TRIG} & \texttt{sub\_EC80D0}, \texttt{sub\_EC80F8}, \texttt{sub\_EBBEDB} \\
\bottomrule
\end{tabular}
\caption{All control bits of the state tracked at address 0x3C52 and its client functions.}
\label{tab:control_bits}
\end{table}

The state machine maintains a running checksum on every byte received
and will mark a complete frame on an IDLE event if the checksum is
correct and the frame length is between 3 and 21 (inclusive), matching
the J1708 specification's \citeproc{ref-saej1708}{23} ``Total message
length, including MID and checksum, shall not exceed 21 characters.''
\footnote{although, according to the spec: \textgreater= 21 bytes is
  permitted when the vehicle is not in motion.} The state machine
injects the length of the received frame minus 1 into the receive buffer
0x3BF5 at the byte preceding the current received frame (e.g., index 0
for the first frame) and increments the frame counter 0x3BF4. Further
frames are stacked in this receive buffer for First-In, First Out (FIFO)
processing outside the interrupt context up to a maximum size of 0x54
bytes (combined data and injected frame lengths).

There are two reads of the SCI2 Data Register Low (DRL), the first (in
the ISR) is a necessary hardware `handshake' to satisfy the peripheral's
requirement of two reads to clear its state, the real data read occurs
within the state machine in \texttt{sub\_F1AFA7}; it reads the data and
catalogs all the errors which are detected by any given UART peripheral
\footnote{This function, like many others encountered, accepts a `device
  index' argument. While other firmwares use non-zero indices for the
  J1708 interface, all target firmwares use index 0 for SCI2/J2497.
  Fortunately, this simplifies structure access and reduces to direct
  offsets, facilitating data xref analysis.} -- but only the data byte
is consumed by the callers, all error information is discarded. This
means that all error detection is deferred to the correct checksum
calculation described previously.

\begin{figure}[htpb]
\includegraphics[width=\columnwidth]{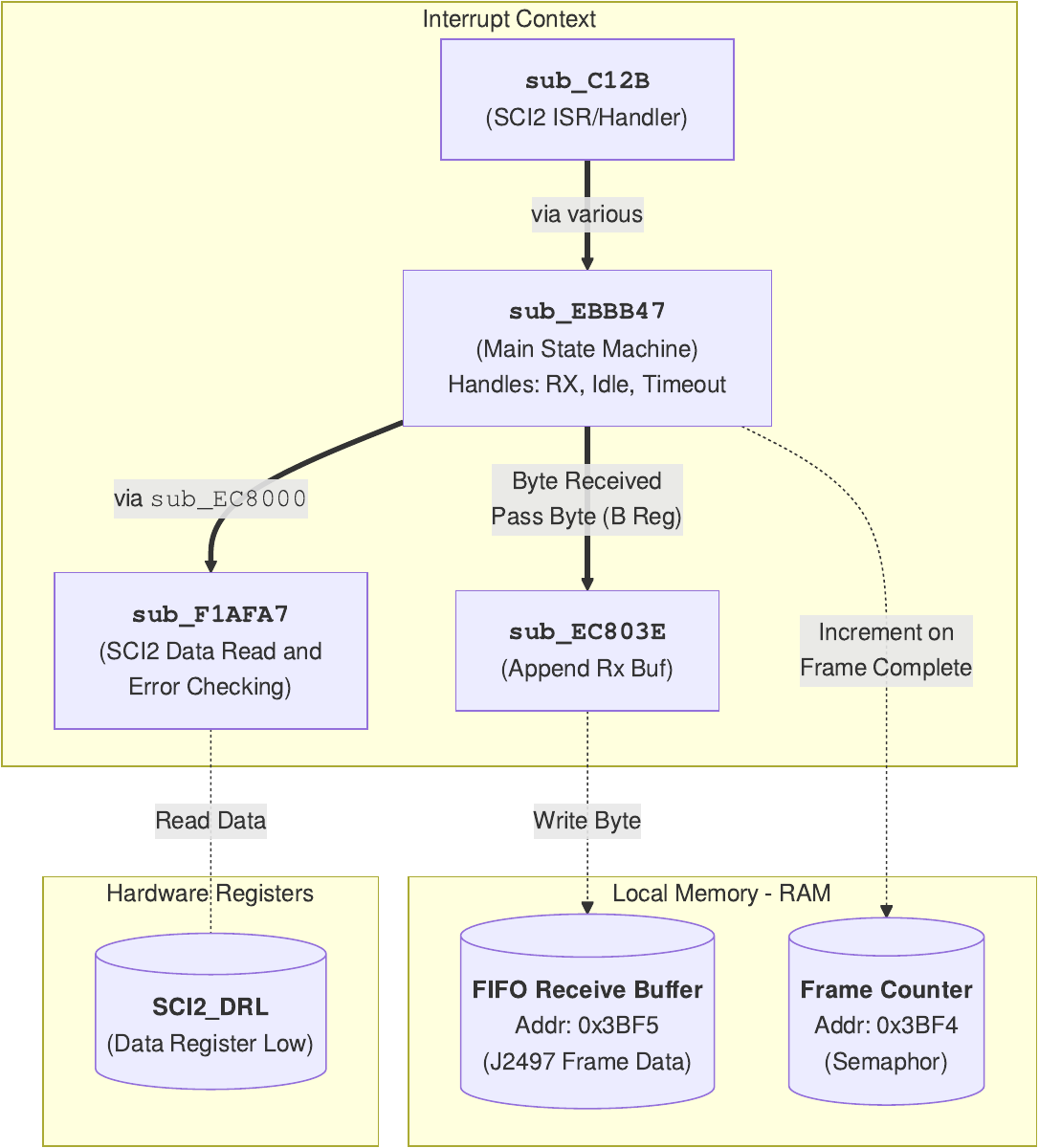}
\caption{Diagram focusing on the dataflow of received J2497 data in the \texttt{2ec80} target. See Figure \ref{fig:sci2_handler_events_buffer} for the details which are omitted here.}
\label{fig:sci2_dataflow}
\end{figure}

\begin{figure*}[hb]
\centering
\includegraphics[width=\textwidth]{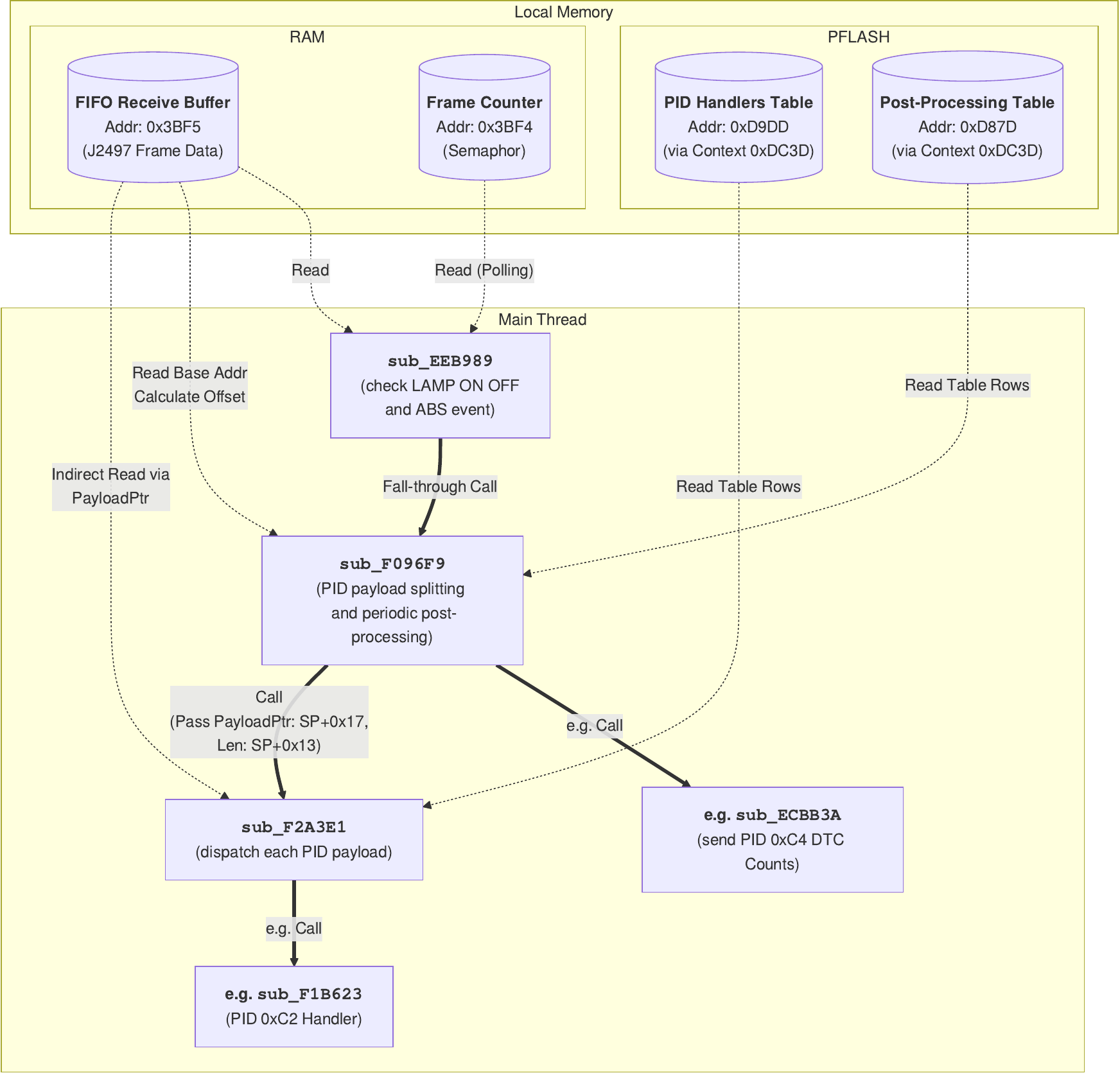}
\caption{Diagram focusing on the dataflow of received J2497 messages in the \texttt{2ec80} target. This diagram picks up where Figure \ref{fig:sci2_dataflow} left off.}
\label{fig:sci2_dataflow2}
\end{figure*}

The J2497 frame reception is realized by a state machine driven by the
events as described above. . As detailed in Section
\ref{limitations-of-dynamic-analysis}, the dataflow is straightforward:
frames are appended to a receive buffer at 0x3BF5, and completion is
signaled via the frame counter at 0x3BF4 (see Figure
\ref{fig:sci2_dataflow}).

\FloatBarrier

\section{Appendix: J1587 PID Processing}\label{j1587-pid-processing}

\begin{table*}[htbp]
\centering
\footnotesize
\rowcolors{2}{gray!10}{white}
\begin{tabular}{@{}l p{0.4\textwidth} p{0.35\textwidth}@{}}
\toprule
J1708 message & PID / PID Request Description & Data Input Size \\
\midrule
\makecell[l]{mm0031 / \\ mm803188} & Request for 0x31 49 ABS Control Status & one byte data (broadcast req) / zero bytes data (unicast req) \\
\makecell[l]{mm003E / mm803E88} & Request for 0x3E 62 Retarder Inhibit Status & one byte data / zero bytes data \\
\makecell[l]{mm0054 / mm805488} & Request for 0x54 84 Road Speed & one byte data / zero bytes data \\
\makecell[l]{mm0097 / mm809788} & Request for 0x97 151 TC Control State & one byte data / zero bytes data \\
\makecell[l]{mm009E / mm809E88} & Request for 0x9E 158 Battery Potential (V), Switched & one byte data / zero bytes data \\
\makecell[l]{mm00A8 / mm80A888} & Request for 0xA8 168 Battery Potential (V) & one byte data / zero bytes data \\
\makecell[l]{mm00C2 / mm80C288} & Request for 0xC2 194 Transmitter System Diagnostic Code and Occurrence Count Table & one byte data / zero bytes data \\
\makecell[l]{mm00C4 / mm80C488} & Request for 0xC4 196 Diagnostic Data/Count Clear Response & one byte data / zero bytes data \\
\makecell[l]{mm00C7 / mm80C788} & Request for 0xC7 199 Traction Control Disable State & one byte data / zero bytes data \\
\makecell[l]{mm00D1 / mm80D188} & Request for 0xD1 209 ABS Control Status, Trailer & one byte data / zero bytes data \\
\makecell[l]{mm00D6 / mm80D688} & Request for 0xD6 214 Vehicle Wheel Speeds & one byte data / zero bytes data \\
\makecell[l]{mm00E9 / mm80E988} & Request for 0xE9 233 Unit Number (Power Unit) & one byte data / zero bytes data \\
\makecell[l]{mm00EA / mm80EA88} & Request for 0xEA 234 Software Identification & one byte data / zero bytes data \\
\makecell[l]{mm00ED / mm80ED88} & Request for 0xED 237 Vehicle Identification Number & one byte data / zero bytes data \\
\makecell[l]{mm00F3 / mm80F388} & Request for 0xF3 243 Component Identification & one byte data / zero bytes data \\
mm2A & 42 Pressure Switch Status & 1 byte data \\
mm46 & 70 Parking Brake Switch Status & 1 byte data \\
mm54 & 84 Road Speed & 1 byte data \\
mm75 & 117 Brake Primary Pressure & 1 byte data \\
mm76 & 118 Brake Secondary Pressure & 1 byte data \\
mmA9 & 169 Cargo Ambient Temperature & 2 bytes data \\
mmB4 & 180 Trailer Weight & 2 bytes data \\
mmC2 & 194 Transmitter System Diagnostic Code and Occurrence Count Table & variable bytes data. byte count, followed by sets of diagnostic data \\
mmC303 & 195 Diagnostic Data Request / Clear & 3 bytes data \\
mmC7 & 199 Traction Control Disable State & variable length data. byte count, state flags, ASCII 'access code' of 0-15 characters selected by the manufacturer ... \\
mmD1 & 209 ABS Control Status, Trailer & variable bytes data, up to 3x5 bytes for 5 trailers byte count, followed by the status bytes \\
mmED & 237 Vehicle Identification Number & variable bytes data. byte count, followed by the VIN ASCII \\
mmF5 & 245 Total Vehicle Distance & 4 bytes data \\
mmF7 & 257 Total Engine Hours & 4 bytes data \\
\makecell[l]{mmFE88C5 / mmFE88C6} & 254 Proprietary DLE & permitted lengths of only 0xC5 and 0xC6 \\
mmFF73 & 115 Trailer Pneumatic Supply Line Pressure & one byte data \\
mmFF7B & 123 Door Status & one byte data \\
\bottomrule
\end{tabular}
\caption{J1587 PIDs supported by the before update firmware. 'mm' refers to any MID value except 0A, 0B, or 57, which are handled earlier in processing.}
\label{tab:j1587_pids}
\end{table*}

The PID (number) is the first byte of the PID payload and this is used
to match a handler in PID handler tables at 0xD9DD (see Table
\ref{tab:pid_handlers}) along with a second-byte match (which can be
0xFF for a wildcard). This table is accessed via a pointer in a `context
structure' (see Table \ref{tab:context}) and the PID handlers row can
set a nested-context structure for any matched PID and second-byte match
(e.g.~this is how 0xFE data link escapes are handled in the firmware).

\begin{table}[htbp]
\centering
\footnotesize
\begin{tabular}{@{} l l p{0.6\columnwidth} @{}}
\toprule
Offset & Size & Description \\
\midrule
0x00 & 1 & Primary identifier for the PID. \\
0x01 & 1 & Secondary identifier (0xFF indicates a wildcard/any). \\
0x02 & 3 & 24-bit pointer to the handler function. \\
0x05 & 1 & Unused byte. \\
0x06 & 2 & 16-bit word. If non-zero, it is treated as a pointer to a nested Context Structure (see Table \ref{tab:context}) for recursive matching and processing of the next bytes in the payload. \\
\bottomrule
\end{tabular}
\caption{PID Handler data structure.}
\label{tab:pid_handlers}
\end{table}

The function, \texttt{sub\_F096F9}, also looks up periodic processing
functions after the PIDs are processed. It uses an assumption of the
rate it will be called by the parent functions and a per-device count to
fire periodic processing calls at rate and phase configured in the table
linked from offset 0xB in the context structure. The call rate is 20Hz
because \texttt{sub\_EEB98} contains an implementation of the J2497 lamp
hysteresis specification in state variables word\_FD126F and flag 0x10
of byte\_4466; the values taken by word\_FD126F only match the
specification if the function is called at 20Hz. It also follows from
this that the non-LAMP PID processing only consumes one J2497 frame per
50ms.

\begin{table}[htbp]
\centering
\footnotesize
\begin{tabular}{@{} l l p{0.6\columnwidth} @{}}
\toprule
Offset & Size & Description \\
\midrule
0x00 & 2 & Used to calculate execution rate (Base 20Hz). Rate = 20 / Divisor. \\
0x02 & 2 & Used to calculate execution phase/offset. Phase = Remainder * 50ms. \\
0x04 & 3 & 24-bit pointer to the handler function. \\
0x07 & 1 & Unused byte. \\
\bottomrule
\end{tabular}
\caption{Post Processing data structure.}
\label{tab:post_handlers}
\end{table}

The context structure \ref{tab:context} holds pointers to the PID
handlers and post-processing tables and total sizes (the post-processing
entries are zero in the nested context structures). It also holds a
limit for the clock-counter in post-processing (also unused in nested
contexts) and a default handler when no PID is matched (implemented as a
null subroutine in the firmwares).

\begin{table}[htbp]
\centering
\footnotesize
\begin{tabular}{@{} l l p{0.6\columnwidth} @{}}
\toprule
Offset & Size & Description \\
\midrule
0x00 & 2 & 16-bit pointer to the PID Table array. See Table \ref{tab:pid_handlers} \\
0x02 & 2 & 16-bit pointer to the Post Processing Table array. See Table \ref{tab:post_handlers} \\
0x04 & 3 & 24-bit function pointer invoked if no PID match is found. \\
0x07 & 1 & Unused byte. \\
0x08 & 2 & 16-bit word (likely a limit for a device counter). \\
0x0A & 1 & Number of entries in the PID Table. \\
0x0B & 1 & Number of entries in the Post Processing Table. \\
\bottomrule
\end{tabular}
\caption{Context (and nested context) data structure.}
\label{tab:context}
\end{table}

The function pointers in these structures were already picked up by the
automated function pointer discovery discussed above. Several pointed at
the same address but there were unique and non-trivial handlers. We
wrote a script to walk the table and name the functions with the
information gleaned from the table context e.g.
``\path{pid_FF_any_default_OR_pid_FE_88_default_OR_pid_80_any_default_OR_pid_00_any_default_OR_nullsub_2_OR_pid_default_handler}
.'' See Appendix \ref{ida-pid-naming} for the script used. The recursive
tables were analyzed in all three before-update firmwares and the PIDs
supported were found to be the same. The supported PIDs are in Table
\ref{tab:j1587_pids}. Note that `supported' here means that the target
ECU will parse incoming payloads of these PIDs; in all cases on J2497
there is no response emitted (whereas there might be on J1708
interface).

\onecolumn

\section*{Appendix: Code Listings}\label{appendix-code-listings}
\addcontentsline{toc}{section}{Appendix: Code Listings}

In this section we provide code listings for the scripts mentioned in
the paper.

These are standalone scripts produced by inlining functions from an
idapython ec80-specific analysis library which runs completely
automatically to create the before and after images of all 3 of the
targets and then performs the qbindiff analysis and produces the plots
in this paper. We are not releasing that script because it contains many
dependencies on the specifics of the EC80 and hence also `leaks'
potentially attacker-beneficial information.

\subsubsection{IDA Pro Python Script: PFLASH and PPAGE
Setup}\label{ida-pflash-setup}

This script sets up the memory map for the S12X processor in IDA Pro,
handling the PFLASH and PPAGE banking.

See file \texttt{ida-pflash-s12x-setup.py} in the paper artifacts.

\subsubsection{IDA Pro Python Script: Concolic
Analysis}\label{ida-concolic-analysis}

This script performs a basic concolic analysis to resolve indirect
calls.

See file \texttt{ida-concolic-s12x-resolve.py} in the paper artifacts.

\subsubsection{IDA Pro Python Script: Jump Table
Detection}\label{ida-jump-table}

This script detects and creates jump tables by parsing the disassembly
and creating manual code xrefs.

See file \texttt{ida-jumptables-s12x-resolve.py} in the paper artifacts.

\subsubsection{IDA Pro Python Script: Function Pointer
Discovery}\label{ida-func-ptr-discovery}

This script scans the firmware image for potential function pointers to
identify additional functions.

See file \texttt{ida-funcptr-s12x-discovery.py} in the paper artifacts.

\subsubsection{IDA Pro Python Script: PID Handler
Naming}\label{ida-pid-naming}

This script walks the PID handler tables and names the handler functions
based on the PID and context.

See file \texttt{ida-pid-ec80-naming.py} in the paper artifacts.

\subsubsection{IDA Pro Python Script: Call Tree
Validation}\label{call-tree-script}

This script validates the call tree to identify deleted functions.

See file \texttt{ida-deletedfn-report.py} in the paper artifacts.

\subsubsection{Fuzzing Script: J1587 PID Fuzzer}\label{fuzzing-script}

This script was used to fuzz the J1587 PID handlers and discover the
vulnerabilities in the 0xC2 and 0xED PID Handlers.

See file \texttt{j1587-pid-fuzzer.py} in the paper artifacts.

\subsubsection{In-motion Test Script: 0xC2 PID
Handler}\label{c2-pid-handler-crash}

This script was used to trigger the 0xC2 vulnerability during in-motion
tests. It emits the test signal on an attached FL2K SDR. It expects the
j2497-keyhole \citeproc{ref-j2497keyhole}{47} project is installed as a
package or that the script runs from the same directory as the project.

See file \texttt{j2497-sdr-badones-transmit.py} in the paper artifacts.

\subsubsection{In-motion Test Script: 0xED PID
Handler}\label{ed-pid-handler-crash}

This script was used to trigger the 0xED vulnerability during in-motion
tests.

Same script as listing \ref{c2-pid-handler-crash}; uncomment the desired
payload.

\subsubsection{HiWave Script: Memory Dump}\label{hiwave-script}

This script uses the HiWave scripting interface to dump memory to a
file. NOTE: you must set the hiwave command window's cache limit to
unlimited before running the script/commands below:

\begin{Shaded}
\begin{Highlighting}[]
\NormalTok{DMM CACHINGOFF}
\NormalTok{DMM WRITEREADBACKOFF}
\NormalTok{HCS12X\_MAP4000 RAM}
\NormalTok{db 0x4000\textquotesingle{}L..0x7FFF\textquotesingle{}L}
\NormalTok{db 0x0f8000\textquotesingle{}G..0x0FBFFF\textquotesingle{}G}
\end{Highlighting}
\end{Shaded}

\section*{Appendix: Disassemblies}\label{appendix-disassemblies}
\addcontentsline{toc}{section}{Appendix: Disassemblies}

REDACTED FOR PREPRINT

\end{document}